\documentclass{article}

\usepackage[utf8]{inputenc} 
\usepackage[T1]{fontenc}    
\usepackage{hyperref}       
\usepackage{url}            
\usepackage{booktabs}       
\usepackage{amsfonts}       
\usepackage{nicefrac}       
\usepackage{microtype}      
\usepackage{xcolor}         
\usepackage{enumitem}
\usepackage{subcaption} 
\usepackage{bm}
\usepackage{graphicx}
\usepackage{amsmath}
\usepackage[a4paper, margin=1in]{geometry}

\begin{document}

\title{
A Distributional Framework for Generative Modeling of Molecular Crystals}

\author{
Michael Kilgour\textsuperscript{1,*} \and
Alex Dong\textsuperscript{1} \and
Mark E. Tuckerman\textsuperscript{1,2,3,4} \and
Jutta Rogal\textsuperscript{5}
}

\date{
\textsuperscript{1}Department of Chemistry, New York University\\
\textsuperscript{2}Courant Institute of Mathematical Sciences, NYU\\
\textsuperscript{3}NYU-ECNU Center for Computational Chemistry at NYU Shanghai\\
\textsuperscript{4}Simons Center for Computational Physical Chemistry at NYU\\
\textsuperscript{5}Initiative for Computational Catalysis, Flatiron Institute\\
\textsuperscript{*}\texttt{michael.kilgour@nyu.edu}
}

\maketitle

\begin{abstract}
Molecular crystals are a highly polymorphic class of materials, with a single molecule commonly crystallizing via multiple packing patterns, making structure and property prediction very challenging.
Crystal structure prediction typically comprises the production of sets of promising candidate structures, each considered in isolation rather than as samples in a thermodynamic distribution.
Likewise, modern generative approaches to this problem, despite naturally sampling distributions of crystals, lack a concrete formulation of the distributions being sampled.
Two components are required to impart meaning to the distributions of crystals generated under such models: a canonical parameterization, and a loss function which equilibrates the generated samples to some target distribution.
We develop such a parameterization, and train energy-based generative flow networks (GFlowNets) to approximate the Boltzmann distribution over crystal structures for target molecules and space groups.
Combined, these components comprise our MXtalGFlow framework for molecular crystal modeling.
Going beyond sampling disconnected sets of low-energy structures, MXtalGFlow yields a thermodynamic distribution over crystal structures.
We sample and analyze distributions of crystals for two molecules, each under two energy functions, a Lennard-Jones potential and the Universal Model for Atoms.
We characterize the local structural basins about the known polymorphs, and identify additional as-yet un-reported packing modes with competitive probabilities to the known experimental structures.
With MXtalGFlow, we illustrate how to define and train a model to sample a thermodynamically meaningful distribution of molecular crystals, and analyze such a distribution to glean useful information.
\end{abstract}

\section{\label{sec:introduction}Introduction}

Computational molecular crystal structure prediction (CSP) is an extremely difficult problem, combining high-dimensional search, quantum chemistry, and thermodynamics.
Generative models promise a dramatic reduction in the computational cost of CSP, allowing practically unlimited sampling at marginal expense after a one-time training investment.
Large sets of crystals can be cheaply generated for downstream refinement and evaluation of promising candidates.
Recent contributions on generative molecular CSP developed and trained molecule-conditioned models on crystal datasets~\cite{jin2025oxtalallatomdiffusionmodel,Subramanian2026,Zeng,vahediahmar2026orgflow,lo2026fast}, though none targeted a thermodynamic distribution and all lacked detailed distributional analysis.

The issue of crystal polymorphism, the propensity of many molecules to crystallize in varied structures~\cite{bernstein2020polymorphism}, requires the ideal crystal generator to produce a diverse distribution of crystals in order to ensure coverage of all such polymorphs.
Unlike Monte Carlo, where one expects in the limit of large sampling to asymptotically cover the full search space, if a generative model misses a given packing mode in training and assigns it a vanishing probability, that packing mode will in practice never be sampled.
This is a major disadvantage for such models in the role of lead generators, unless serious steps are taken to guarantee coverage of all packing modes.

In addition, since generative models are inherently stochastic and distributional objects, it is worthwhile to explore, not just the diversity of such distributions, but also their physical content.
It has been proposed, for example, that training a generative model on experimental datasets provides kinetic and thermodynamic information about the true physical distributions of crystals that may be missed under a pure energy minimization strategy~\cite{jin2025oxtalallatomdiffusionmodel}.
While this is plausible in principle, the distributions of crystals in the major experimental molecular crystal databases are highly biased, with crystals grown via a vast range of largely unrecorded experimental conditions and crystallization strategies.

Inorganic and organic molecular crystal structure prediction tools are typically evaluated by their performance on sample quality and diversity metrics~\cite{metni2026generative}; one wants to know whether the samples are energetically reasonable and whether the model samples multiple modes of the distribution.
In data-driven workflows, `diversity' is defined with respect to some reference distribution, such as the training data or other generated distributions.
While the quantification of this type of diversity has advanced in inorganic CSP to include sophisticated distributional analysis~\cite{klipfel2024vector}, diversity is still defined intrinsically against a reference distribution.
Such distributions, whether representing an experimental dataset or generated by a data-driven model, are not guaranteed to cover all the low-energy crystal modes of a given molecule. 
Indeed, without an extrinsic, canonical map of the search space, no such reference distribution can be confidently asserted to include all low-energy modes.
A systematic search requires a map.

Given a parameterization rigorously derived from physical degrees of freedom, it becomes possible for a generative model to learn a physical distribution.
This is indeed the insight and goal of the Boltzmann Generator literature~\cite{noe2019boltzmann}, which leverages the properties of normalizing flow models to enable modeling of the Boltzmann distribution over physical systems.
The broader literature on energy-based models, and specifically generative flow networks~\cite{bengio2023gflownet} (GFlowNets or GFNs), addresses the same problem, though with different architectures and training modalities. 
Originally developed for the discrete generation of directed acyclic graphs~\cite{bengio2021flow}, GFlowNets have evolved to enable the generation of wide ranges of statistical objects, with numerous losses and training protocols~\cite{lahlou2023theory,jain2022biological,zhang2022unifying}.
It is now possible with such tools to train models that approximate the Boltzmann distribution defined by an energy function and a target temperature, even for very rugged and high-dimensional spaces such as in molecular crystal packing.
Practically, such a model would sample all the low-energy configurations for a given molecule, with probability approximately proportional to their Boltzmann weights. 

In a typical generative CSP workflow, leads may be sampled with a generative model, and an energy function might be used for downstream refinement.
By contrast, in energy-based modeling, the energy function itself in combination with the molecule and crystal parameterization \textit{defines} the distribution of crystals sampled by the model.
This provides a much richer view of the thermodynamic impacts of different energetic assumptions, with, e.g., a deep but very narrow energetic basin that appears energetically plausible but is thermodynamically down-weighted due to its tiny volume.
Far from merely re-ranking or re-optimizing particular structures with different energy functions, the physical distribution --- the size, shape, and relative locations of energetic basins --- may significantly shift depending on the selected potential.

In this work, we propose a workflow that opens the door to both the generation and rigorous analysis of distributions of molecular crystals.
We first develop a physically grounded parameterization for molecular crystals, which enables the training of energy-based models with meaningful sample probabilities.
Due to its focus on the asymmetric unit, this parameterization generalizes to all molecules and space groups. 
Then, we build and train a hybrid data- and energy-based GFlowNet model in this parameterization, for a given molecule in a particular space group.
In the convergent limit, GFlowNets of this type sample the Boltzmann distribution over crystal parameters at a target temperature.
These models are known to be very challenging to train on rugged, high-dimensional search spaces; we develop a protocol to ensure stable training without mode collapse or overfitting.
On top of concrete notions of sample diversity, our GFlowNet models yield rich distributional information, including sample probabilities and basin geometries, as we demonstrate on two molecular crystal systems with known experimental structures.
We also illustrate how crystal distributions may vary significantly by comparing crystal landscapes under the Universal Model for Atoms (UMA)~\cite{wood2025family,gharakhanyan2025open} and a Lennard-Jones potential.

\section{\label{sec:Results}Results}

\subsection{\label{sec:Parameterization} Canonical Map}
To generate and analyze distributions of crystal structures, we require a physically grounded crystal parameterization scheme.
Such a scheme should cover the space of molecular crystal configurations without unit cell degeneracy, i.e., identical crystal lattices with different unit vectors should be prohibited.
This parameterization can then be transformed into a regularized `latent' basis with appropriate bounds and periodicities for the model to work in.

\begin{figure}
\centering
\includegraphics[width=0.85\textwidth]{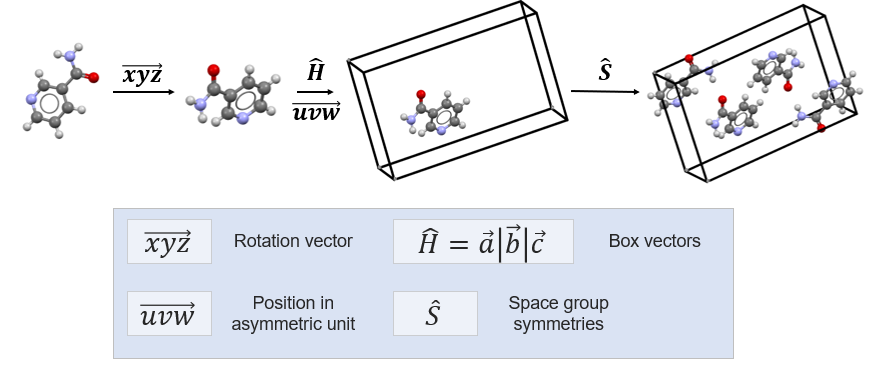}
\caption{\label{fig:workflow} Illustration of our crystal parameterization scheme and crystal building workflow. 
In energy-based training, the GFN model proposes latent parameters $\mathcal{\tilde{C}}$, which are transformed to physical crystal parameters $\mathcal{C}$, and combined with the conformer and space group symmetry to instantiate a crystal unit cell.
The unit cell is then passed to the energy function of choice for reward evaluation on-the-fly.}
\end{figure}
We adopt and improve the parameterization scheme from~\cite{kilgour2025multi, kilgour2025mxtaltoolstoolkitmachinelearning}, defining a low-dimensional representation for molecular crystals, $\mathcal{C}$, as diagrammed in Figure~\ref{fig:workflow}.
Space groups above $P\bar{1}$, $SG\in(3,230)$, all have unit cell vectors with one or more fixed lengths or angles, and so are effectively lower dimensional in the box vectors.
The same parameterization scheme applies in all space groups, with constant dimensions, e.g., $\alpha,\beta,\gamma=\frac{\pi}{2}$ and $a=b=c$ in cubic space groups.
We apply crystal-system-specific restrictions to ensure unit cell box uniqueness, following standard cell standardization schemes~\cite{togo2024spglib}.
For simplicity, in this study, we consider $Z'=1$ crystals formed of rigid molecules.
Our latent transform, from $\mathcal{C}\to\mathcal{L}\in[-1,1]^{12}$, is not volume preserving with respect to the molecular degrees of freedom, and we derive the appropriate Jacobian corrections in the supporting material (SM).

\subsection{\label{sec:GFN}GFlowNet}
For our generative model, we select the GFlowNet family, and specifically specifically the `diffusion sampler' flavor, as developed in~\cite{sendera2024improved}.
Due to the difficulty of converging such models on challenging problems such as molecular crystals, we developed a novel training protocol designed to guarantee mode coverage and model stability.
This type of GFlowNet is a neural stochastic differential equation (SDE), where the drift and variance are learned neural network functions.
What makes an SDE a GFlowNet is the loss function and training protocol: the self-consistent trajectory balance (TB) loss, trainable both on- and off-policy.
The TB loss balances the probabilities of forward and backward policy models, a reward function, and a learned scalar, $\log{Z_\theta}$ (see all details in Section~\ref{sec:GFN_methods} and the SM).
A major advantage of this model is its amenability to both energy-based on-policy training, and stable off-policy training, allowing for thermalization and exploration while guarding against mode collapse. 

GFlowNets are trained on a reward function, for physical systems usually given as a Boltzmann factor,
\begin{equation}
    R(\bm{x})=e^{-\beta E(\bm{x})},
\end{equation}
for $\beta=\frac{1}{k_BT}$ the inverse temperature and $E(\bm{x})$ an energy function evaluated on a crystal parameterized by $\mathcal{C}=\bm{x}$. 
It has been shown~\cite{lahlou2023theory, bengio2023gflownet} that when the model has full support over the true distribution, and the trajectory balance loss converges to zero, the model samples from the Boltzmann distribution defined by the reward function,
\begin{equation}
    p_F^T(\bm{x})=\frac{R(\bm{x})}{Z},
\end{equation}
with $p_F^T$ the terminal distribution over samples $\bm{x}$ from trajectories $\tau\sim p_F$, and
\begin{equation}
Z = \int\;{\rm d}{\bm x}\;R({\bm x}),
\end{equation}
is the partition function.
Provided the TB condition holds exactly for every trajectory, the scalar $Z_{\theta}$ equals the true partition function $Z$.
Perfect convergence is rarely observed except on simple problems, and we analyze the quality of our models' convergence in the SM.

A known issue in the GFlowNet literature is balancing coverage of all high reward modes of the distribution with efficient convergence, with many recent works attacking different aspects of the problem~\cite{Kwon2026,Morozov2026,Niu2026,Yu2026,Chen2026, Malek2025, Rector-Brooks2023}. 
Achieving satisfactory convergence over all packing modes is a major engineering challenge for molecular crystals, which are high-dimensional and have very wide variation in sample energies, on the scale of hundreds to thousands of $kJ/mol$.
We designed a reliable workflow, outlined in Section~\ref{sec:protocol}, for converging GFlowNets on such complex and difficult potential energy surfaces, enforcing coverage and insuring against mode collapse and buffer overfitting.
A key strategic choice is splitting the mode discovery task from the equilibration of the target distribution.
GFlowNets are a uniquely powerful tool for the Boltzmann weighting of samples across complex multimodal distributions, but classical CSP tools are currently more efficient at systematic mode discovery.

\subsection{\label{sec:analysis}Landscape Analysis}
Sampling from a trained MXtalGFlow model yields a distribution of crystal parameters, $\{\bm{x}\}$, for a given molecule and space group.
This distribution contains useful thermodynamic information: principally, the locations and sample probabilities of various crystal packing modes.
The carving of distributions of crystals into neatly defined clusters, and the assignment of clusters to representative polymorph structures to which we give the titles `Form I', `Form II', etc., is itself a complex and difficult modelling task, as has been explored in recent studies~\cite{butler2023reducing,yang2021exploration,yang2022global}.

We propose a simple yet robust analysis protocol based on the per-sample probability density $P(\bm{x})$,
\begin{equation}
\label{eq:pofx}
    P(\bm{x_i}) =\frac{1}{c} \sum_{j\in d_{cut}} e^{-\frac{d_{ij}^2}{2\sigma^2}},
\end{equation}
for $c$ a global normalization factor, $d_{ij}$ the radial distribution function earth mover's distance (RDF EMD, defined in SM) between the crystal parameterized by $\bm{x_i}$ and neighbors $\bm{x_j}$ within a cutoff radius $d_{cut}$, and the bandwidth $\sigma=d_{cut}/3$.
$d_{cut}$ is taken as the 15\% quantile of the all-to-all distance matrix over the full distribution of crystals.
From our experiments, $d_{cut}$ of this magnitude results in a $P(\bm{x})$ that aggregates and smooths probability mass roughly on the length scales of packing modes.

The RDF EMD is a radial density fingerprint distance, similar to the root mean squared atomic distances evaluated using the COMPACK algorithm~\cite{chisholm2005compack}, with the advantages that it is a smooth metric and much faster to compute.
Thus, the $P(\bm{x})$ defined by this distance metric is larger for $\bm{x}$ where there are many physically similar crystal samples in the generated distribution, and lower where the distribution is sparse.

Traditionally one identifies distinct polymorphic structures by local minimization of the potential energy, and deletion of highly similar structures (de-duplication).
With $P(\bm{x})$ in hand, we can instead search for maxima on the probability density field, and find the set of maximally likely representative structures.
Our optimization procedure is discrete hill-climbing on the graph defined by the distribution of samples from MXtalGFlow.
A graph is defined as nodes $\{\bm{x}\}$ with feature $P(\bm{x})$ and bi-directional edges drawn between nodes with RDF EMD less than a maximum step size, $d_{step}$.
All nodes follow steepest ascent on the graph until convergence, resulting in a set of local maxima, $\{\bm{x}_\alpha\}$.
For small $d_{step}$ this procedure results in large numbers of insufficiently distinct maxima, and so we scan over increasing $d_{step}$ until reaching a stable plateau in the number of maxima.

To aid in visualization, we assign samples to basins centered on each sample in the set of maxima, $\{\bm{x}_\alpha\}$, on the basis of physical similarity.
Basin membership is assigned according to Gaussian weights $w_{i,\alpha} = \exp\!\left(-\frac{d_{i,\alpha}^2}{2\sigma^2}\right)$, for each local probability maximum $\alpha$, where $d_{i,\alpha}$ is the RDF EMD between point $\bm{x_i}$ and $\bm{x_\alpha}$, and the same $\sigma$ as above.
Sample $i$ is assigned to the basin defined by $\alpha$ if the normalized basin weight $\bar{w}_{i,\alpha}=\frac{w_{i,\alpha}}{\sum_\beta w_{i,\beta}}>0.8$, indicating a strong preference for just one basin, and otherwise unassigned.

We emphasize that our $P(\bm{x})$ is not intended as a proxy or replacement for quantitative free energy estimates along the lines of the (quasi-)harmonic vibrational free energy~\cite{hoja2017first}.
$P(\bm{x})$ is a measure of relative basin probability, suitable for our task of identifying and characterizing structural basins with high probability density across the full sampled distribution, and is not directly comparable to existing free energy methods.

\subsection{Experiments}
\begin{figure}
\centering
\includegraphics[width=0.5\textwidth]{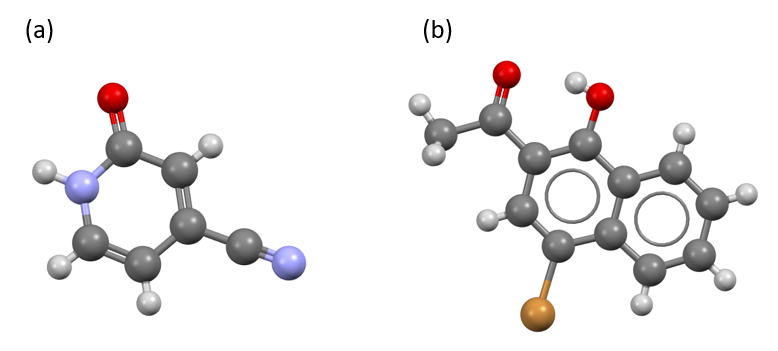}
\caption{\label{fig:molecules} Diagrams of MIPCAS and NEHZOR conformers in (a) and (b).}
\end{figure}
We undertook studies of two illustrative polymorphic landscapes for different molecules and space groups.
The molecule 2-oxo-1,2-dihydropyridine-4-carbonitrile, which we will refer to by its CSD refcode, MIPCAS, is shown in Figure~\ref{fig:molecules} (a).
MIPCAS is a rigid molecule with significant hydrogen bonding character. 
It has one known experimental polymorph in the $P\bar{1}$ space group, which makes it a good test case for our method, as its crystal distribution requires the full 12 dimensions of our parameterization.
We also study the molecule 1-(4-bromo-1-hydroxynaphthalen-2-yl)ethan-1-one, (CSD refcode NEHZOR), shown in panel (b).
NEHZOR is also mostly rigid, and its aryl bromide, as well as hydrogen bond donors and acceptors give it non-trivial energetics.
There are two experimentally solved crystal polymorphs in the $P2_1/c$ space group, NEHZOR and NEHZOR01. 
In this work, we index the structures as Polymorph I and Polymorph II for NEHZOR and NEHZOR01, respectively.
The experimental molecular conformers in both polymorphs are almost identical; we treat the molecule as a single rigid object for all experiments.

These two molecules were selected in order to demonstrate the scope and flexibility of MXtalGFlow in common space groups and to illustrate clearly the influence of different energy functions on their polymorphic landscapes.
The presence of functional groups such as aryl bromide and nitrile, which may be treated very differently by different potential energy functions, significantly influences the crystal distributions learned by our GFN models.

GFlowNets were trained on reward functions of the form $R(x)=\exp{(-\frac{E(\bm{x})}{k_{\rm B}T})}$, with $E(\bm{x})$ the energy of a given crystal structure and $k_{\rm B}T=2.5$\ kJ/mol, or roughly room temperature.
The energy function is the sum of an intermolecular energy function, either Lennard-Jones or UMA, and additional terms for training stabilization, cell uniqueness, and Jacobian corrections, all given in the SM.
After training, 10k crystal samples were generated by each model for the analyses below. 

\subsection{\label{sec:landscapes}Crystal Landscapes}

\begin{figure}[!ht]
\centering
\includegraphics[trim=0 340 0 0, clip,width=\textwidth]{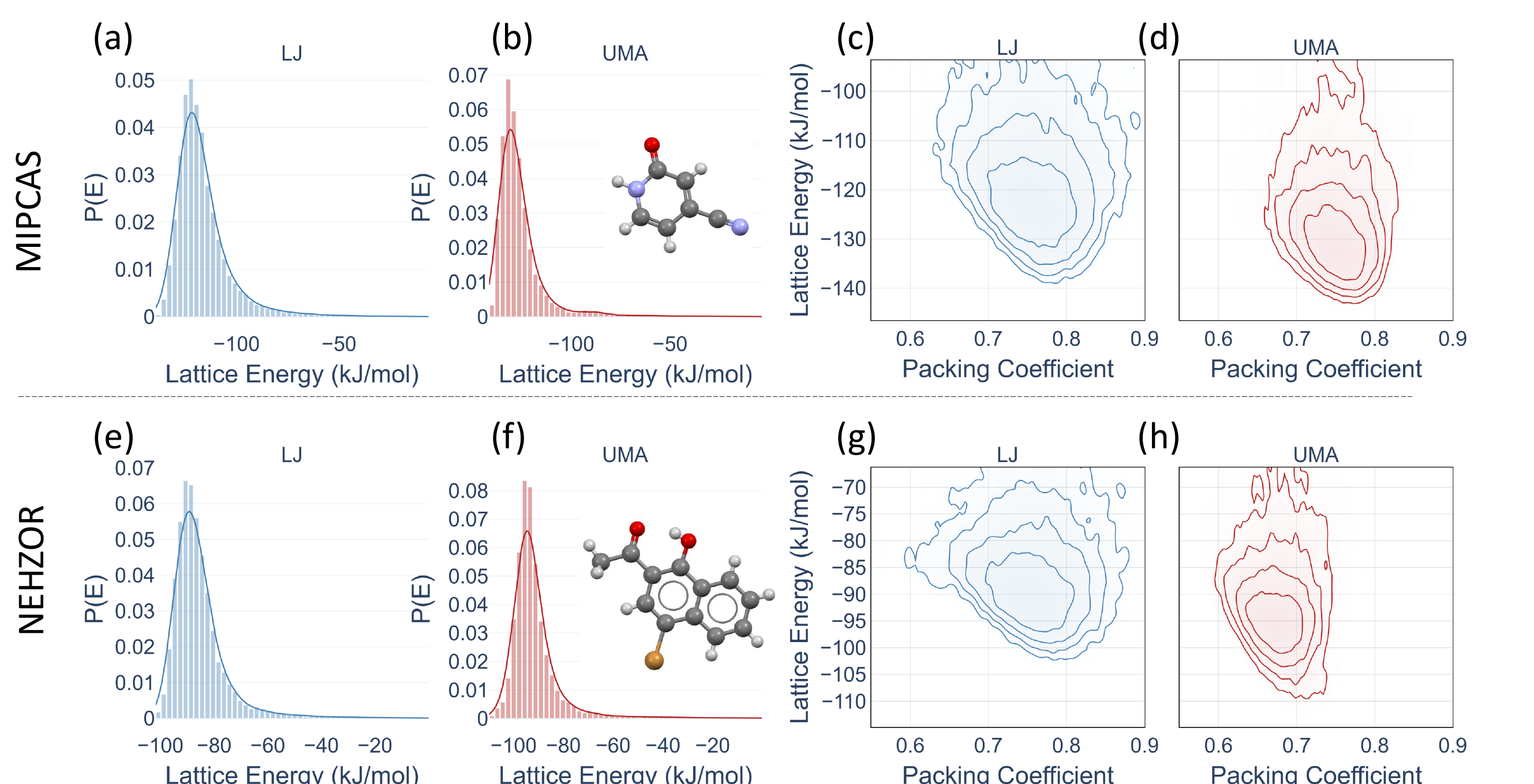}
\caption{\label{fig:energies} Summary energy and density statistics for distributions of crystals sampled from trained MXtalGFlow models. 
Marginal distributions over energy in panels (a-b), (e-f). Energy-density contour plots for samples from LJ and UMA models in panels (c-d) and (g-h).
The 1\% quantile of highest energy samples are omitted for visual clarity.}
\end{figure}
In Figure~\ref{fig:energies}, we show lattice energy distributions and lattice energy vs. density contour diagrams for the distributions sampled from trained models on the two molecules and two energy functions.
As expected, the lattice energies peak near the bottom of the energy range (panels a-b, e-f).
We also observe the typical energy vs. density frontier (panels c-d, g-h), where higher densities, up to a point, admit lower energies.
For both molecules, we observe that the LJ distribution has a wider range of densities and, especially for NEHZOR, higher maximum densities.

The Boltzmann distribution of energies at finite temperature takes the form 
\begin{equation}
\label{eq:pofe}
    P(E)\propto g(E)e^{-\beta E},
\end{equation}
with the density of states $g(E)$ arising from the projection from state to energy space.
The form of $g(E)$ is system specific and generally unknown, although the behavior in Figure~\ref{fig:energies} is qualitatively consistent with our expectations; the density of states suppresses the probability density near the energetic minimum, and the Boltzmann factor suppresses the probability density at higher energies.

GFN models of this type tend to over-weight higher energy states systematically, as we observe in the long tail of the energy distributions.
This is an artifact of imperfect convergence and can be partially ameliorated with longer training time or larger policy models, though larger models often introduce significant training instability.
We omit manual sample reweighting~\cite{noe2019boltzmann} and, instead, present the learned distributions as they are, accepting these tails as a consequence of imperfect convergence. 
We are focused here on the more important low-energy region, where the GFN fit is much tighter, as we show in the convergence analysis in the SM.

\begin{figure}[!ht]
\centering
\includegraphics[trim=0 180 0 240, clip, width=\textwidth]{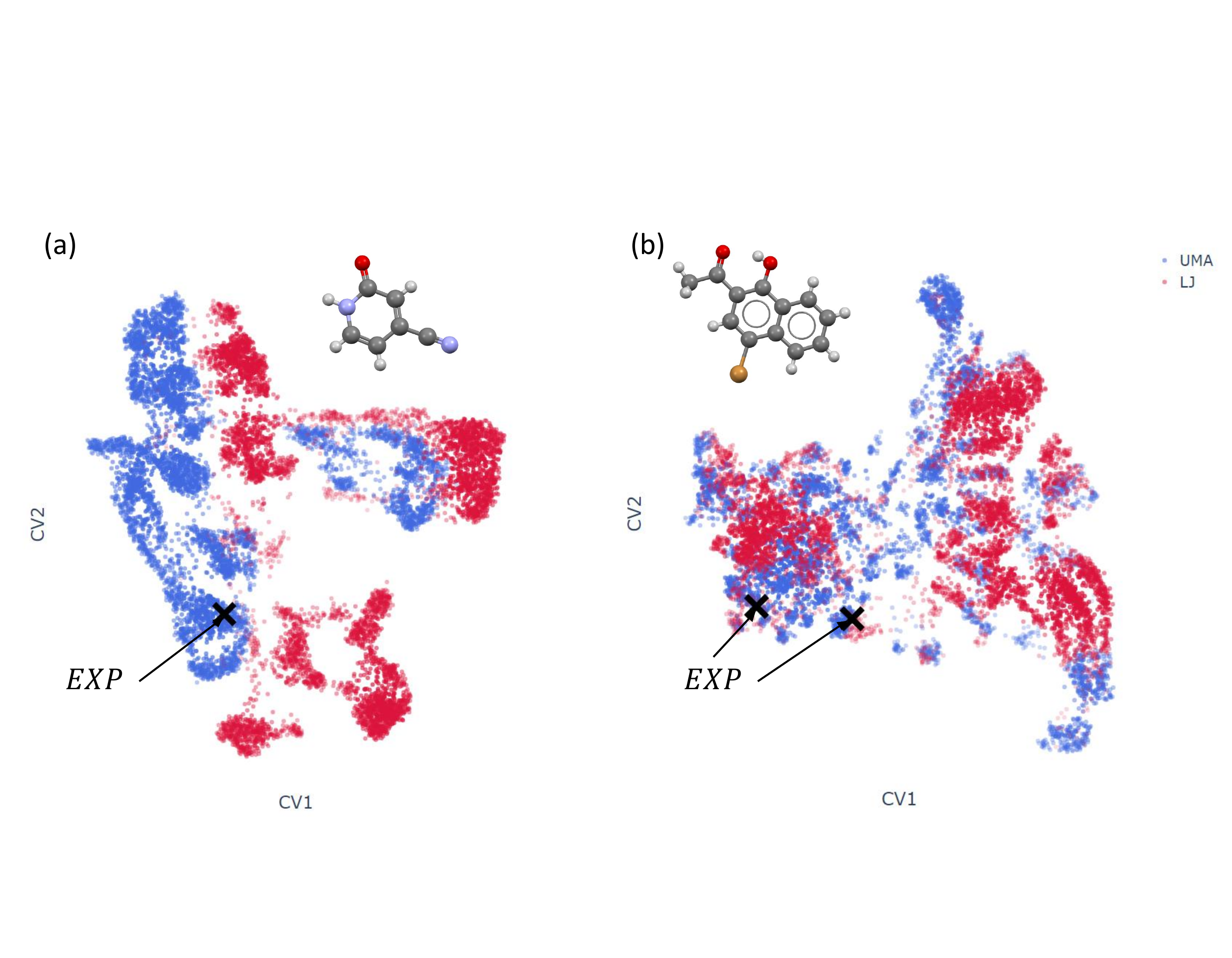}
\caption{\label{fig:lj_uma_embeds} UMAP embeddings of RDF EMD matrices for 5k samples each from LJ and UMA GFlowNet models for $P\bar{1}$ MIPCAS and $P2_1/c$ NEHZOR.
Known experimental polymorphs are marked with a large X.}
\end{figure}

MXtalGFlow enables the generation and comparison of distributions of crystals under different energy functions. 
While the raw  distributions over the latent parameters are useful for quantifying coverage of the search space, they are difficult to interpret thermodynamically.
This is because while short euclidean distances in the latent distribution correspond to physically similar crystals, the converse is often untrue, with distant samples often being physically similar or identical (up to global translation, rotation) due to the action of the space group normalizer on our asymmetric unit parameterization.
We report these distributions in the SM, and here instead employ the RDF EMD as a physically reliable metric for the analyses of crystal distributions.
In Figure~\ref{fig:lj_uma_embeds}, we show two-dimensional UMAP~\cite{mcinnes2018umap} embeddings of the pairwise RDF EMD matrix, with 5k samples from the LJ and UMA GFlowNet models for both molecules.

The UMAP algorithm attempts to cluster points according to a distance metric, in our case placing physically similar crystals close together and dissimilar crystals far apart.
We observe in panel (a) that the embeddings of crystals for $P\bar{1}$ MIPCAS under LJ and UMA potentials are almost completely disjoint.
Physically, this arises simply from the lack of hydrogen bonding under the LJ potential.
The UMA distribution, and the experimental polymorph, on the other hand, are dominated by hydrogen bonds, and so their RDF fingerprints are easy to distinguish from the LJ distribution.

The embedding of $P2_1/c$ NEHZOR crystals in panel (b) is more nuanced. 
The LJ and UMA landscapes are overall less dominated by a few large lobes, and have significantly more overlap.
This is because the packing of NEHZOR is less controlled by hydrogen bonding relative to sterics, which are more adequately captured by the LJ energy. 
Nevertheless the two distributions are still mostly non-overlapping, and as we analyze in detail in the next section, the experimentally identified polymorphs are both in regions with high local UMA probability densities and very low LJ probability density.

\subsection{Basin Analysis}
We map the polymorphic landscapes of $P\bar{1}$ MIPCAS and $P2_1/c$ NEHZOR sampled from the GFlowNet models trained on UMA, due to its much higher physical fidelity compared to Lennard-Jones.
The crystal distributions are analyzed on the basis of physical similarity, using the RDF EMD, probability density estimation, and neighborhood assignment method defined in Section~\ref{sec:analysis}. 
Recall that the probability densities we report are smoothed and aggregated over samples within a cutoff radius $d_{cut}$.
Therefore, the $P(\bm{x})$ we assign to a given state represents its structural neighborhood rather than the point-wise density exactly at $\bm{x}$.
As above, the UMAP embeddings are qualitative representations used to improve interpretability, and are not used in the quantitative analysis.

Prior to performing the analysis, we filter crystal samples with energies greater than $15\ k_{\rm B}T$ above the minimum, on average 5\% of the samples.
Such unphysically high-energy samples arise from imperfect convergence, as discussed in Section~\ref{sec:landscapes} and the SM.
Very high-energy crystals are usually far in RDF space from the low-energy region, therefore this pruning has a small impact on our analysis.

\begin{figure}[!ht]
\centering
\includegraphics[trim=0 50 0 5, clip,width=\textwidth]{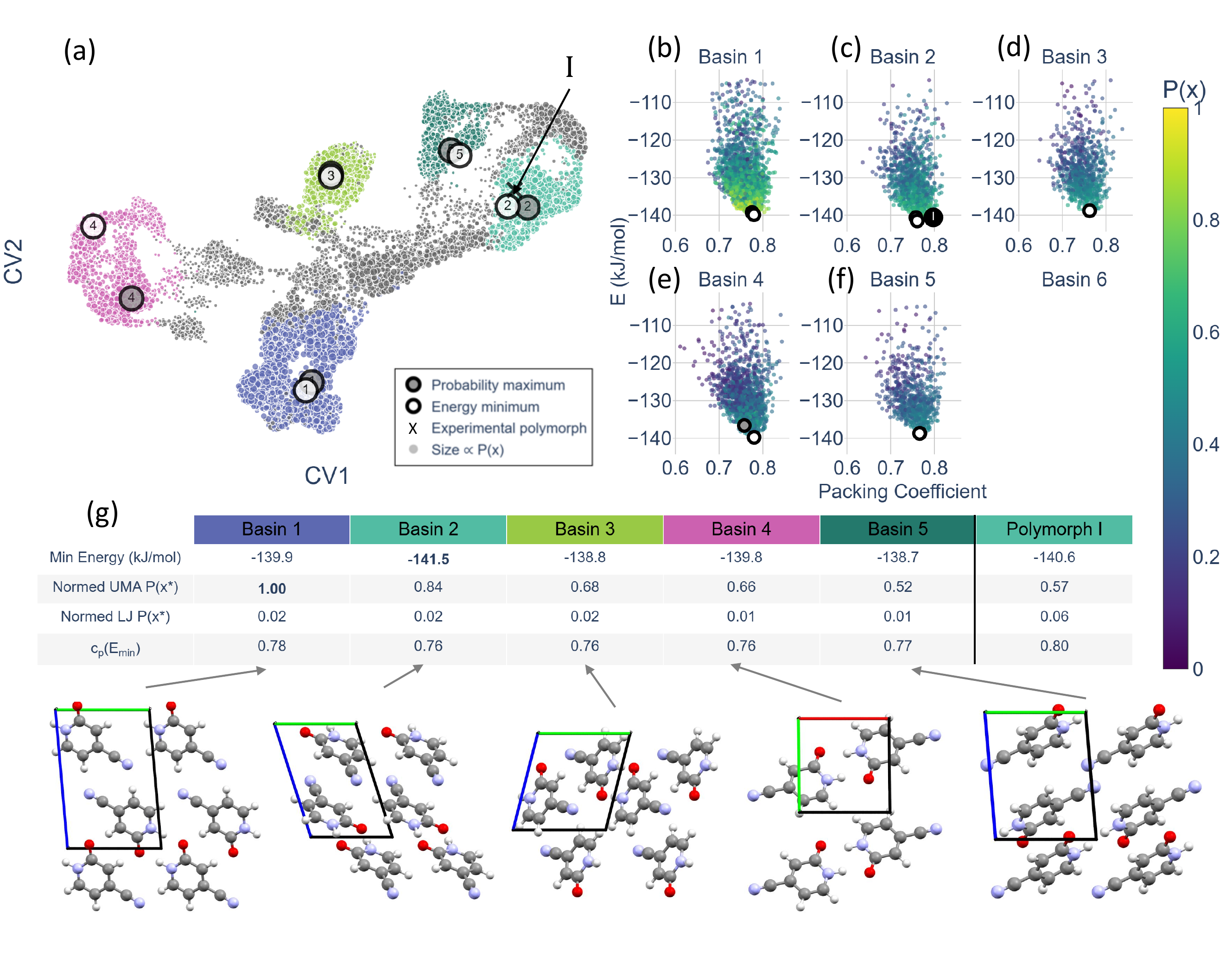}
\caption{\label{fig:mipcas_embedding} 
Landscape analysis for the learned $P\bar{1}$ MIPCAS distribution under UMA. 
(a) A UMAP embedding of the RDF EMD matrix, with colors identifying assignments to the top basins. 
(b-f) Energy vs. density plots for each basin, with colors normalized against the probability maximum. 
(g) Summary statistics for each basin, with the normalized probability densities evaluated at $x^*$, the basin maxima under UMA, and for the experimental polymorph. 
Images of maximal probability structures in each basin are shown below. 
The reported maximum basin probability is normalized against the maximum of all basins.}
\end{figure}
In Figure~\ref{fig:mipcas_embedding}, we analyze the learned $P\bar{1}$ landscape for MIPCAS under UMA. 
Panel (a) shows the UMAP embedding over RDF distances, where we label several distinct probability density maxima and their surrounding basins.
Though only a qualitative clue, it is encouraging that the shapes of the dense clusters generated by UMAP roughly track our similarity-based basin identification scheme. 
Basins are connected variously by thin or thick filaments in the embedding space, indicating a lower or higher degree of inter-basin connectivity.
In Panel (b-f), we report energy vs density distributions within each basin, with the globally normalized probability density indicated by the color.
As expected, within-basin probabilities are higher at lower energies.

The experimental polymorph is found in basin 2, which has the lowest minimum potential energy of all basins, with the energetic minimum quite close to the polymorph in RDF space, as indicated in the UMAP embedding in (a).
COMPACK analysis of several samples nearest to the experimental polymorph in RDF distance yields several 20/20 molecular cluster matches with RMSDs below 0.3\AA, confirming Polymorph I is securely nested inside the assigned basin.
This is unsurprising, as COMPACK distances tend to correlate well with the RDF EMD, especially at short distances.
We also see in panel (c) that the experimental polymorph is among the lowest energy samples in basin 2, and among the densest. 

We see in the summary table, panel (g), that compared to basin 2, all other basins have marginally higher minimum energies, and the maximum probability density in basin 1 is slightly higher than basin 2.
According to Eq.~\ref{eq:pofe}, this indicates that basin 1 has a higher configurational density of states than basin 2, offsetting its lower Boltzmann weight.
Assuming our model and the UMA potential are physically faithful, our analysis therefore predicts that basin 1 is slightly thermodynamically preferred over the known experimental structure in basin 2.

Panel (g) also includes the LJ probability density evaluated at the UMA probability optimum, $x^*$, for each basin.
Given our observation in Figure~\ref{fig:lj_uma_embeds} that the physical manifolds for LJ and UMA are almost completely disconnected, it is logical that the probability densities for all UMA-derived basins and the experimental polymorph under LJ are small.
While LJ vs UMA is perhaps an extreme case of comparing different energy functions, the point stands that distributions of crystals under different potentials yield significantly different downstream thermodynamic observables.

\begin{figure}[!ht]
\centering
\includegraphics[trim=0 0 0 5, clip, width=\textwidth]{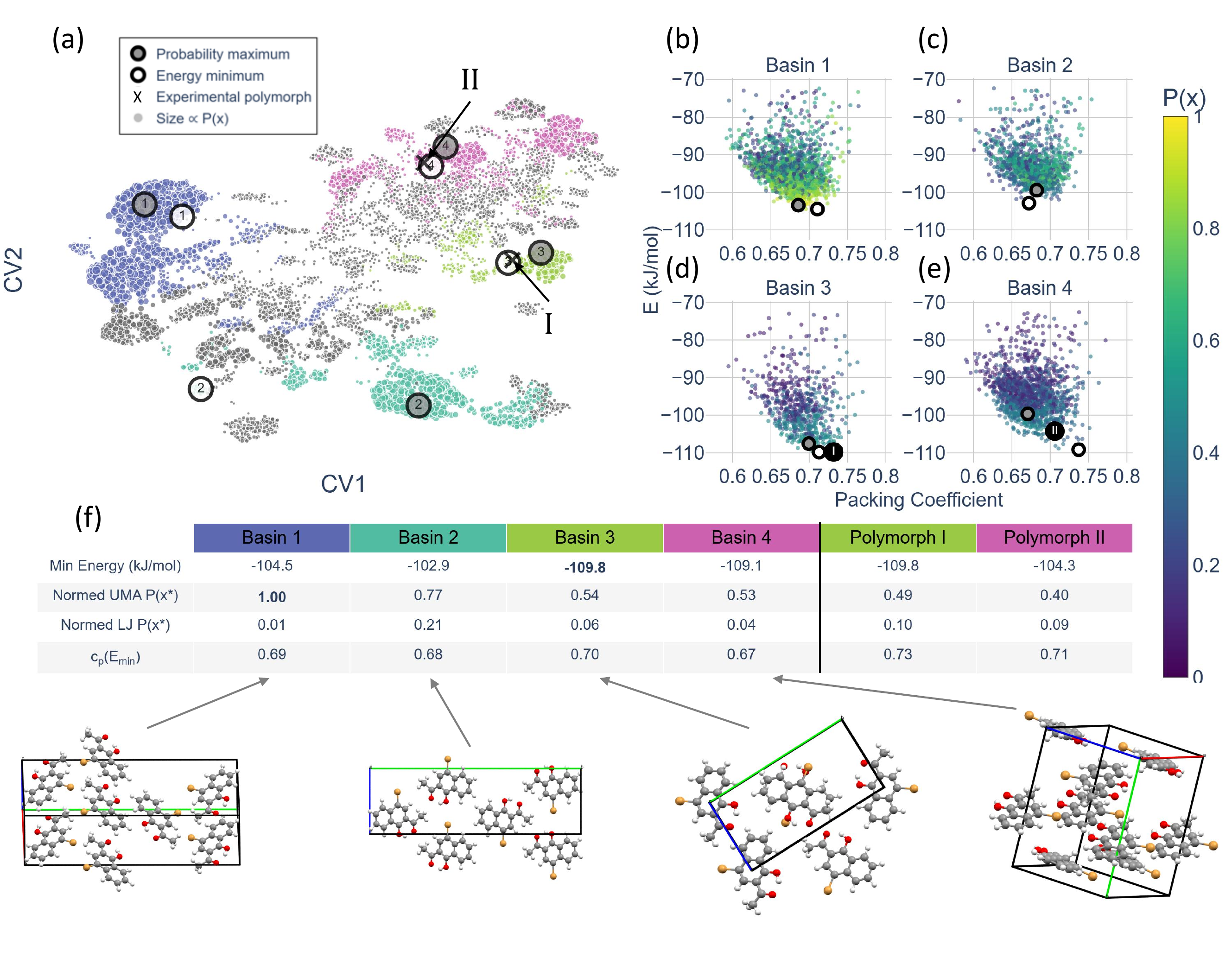}
\caption{\label{fig:nehzor_embedding} 
Landscape analysis for the learned $P2_1/c$ NEHZOR distribution under UMA. 
(a) A UMAP embedding of the RDF EMD matrix, with colors identifying assignments to the top basins. 
(b-e) Energy vs. density plots for each basin, with colors normalized against the probability maximum. 
(f) Summary statistics for each basin, with the normalized probability densities evaluated at $x^*$, the basin maxima under UMA, and for the experimental polymorphs.  
Images of maximal probability structures in each basin are shown below. The reported maximum basin probability is normalized against the maximum of all basins.}
\end{figure}
The $P2_1/c$ NEHZOR landscape, shown in Figure~\ref{fig:nehzor_embedding}, is rather rich, with two known polymorphs and several structural basins.
We again see in panel (a) UMAP clusters correlate with probability basins, though the overall landscape is characterized by a larger number of smaller clusters, rather than just a few large ones.
Basins 1 and 2 dominate the largest and most visually distinct regions, whereas basins 3 and 4 share a more structurally mixed region.

Both experimental polymorphs are again found in the lowest energy basins, with Polymorph I very close to the energy minimum for basin 3, and Polymorph II slightly higher than the minimum of basin 4.
As with the previous landscape, a number of samples nearby to the experimental polymorphs in RDF space are 20/20 COMPACK matches, with RMSDs below 0.3\AA, confirming the polymorphs' assignments to their respective basins.

Basins 1 and 2, despite having higher mean and minimum energies, have notably higher probability densities than 3 and 4, where the known polymorphs are found.
As with $P\bar{1}$ MIPCAS, this necessitates a higher configurational density of states in these basins, and to an even greater degree.
Assuming again that UMA is a suitable potential for this system, and granting our modeling assumptions, our protocol predicts basins 1 and 2 are thermodynamically preferred compared to the known polymorphs at room temperature.

\section{\label{sec:summary}Summary and Conclusions}
We present the MXtalGFlow framework for the modeling of molecular crystals on the level of distributions of crystal structures, resting on three conceptual pillars: 1) a canonical parameterization, 2) an energy-based generative model, and 3) distributional analysis based on physical similarity. 
To realize this, we developed a parameterization and latent transform for molecular crystals, suitable for use in a generative model and tethered to the physical degrees of freedom.
Our parameterization defines a `map' of the available search space within a given space group, which can be used to quantify coverage of all packing modes, as well as to define a physically grounded thermodynamic distribution. 
With this, we evolve generative molecular CSP from the sampling of disconnected sets of crystal structures to the modeling of a physical distribution over a canonicalized search space.

We built a training protocol for GFlowNet diffusion samplers engineered so as to cover the search space while avoiding common pathologies even on rugged, high-dimensional energy landscapes.
By pre-constructing a dataset that defines the support of the distribution, we separate the mode discovery task from the equilibration task, with the latter being the specific and unique capability afforded by GFlowNets. 
By training in three controlled stages: 1) expand model support, 2) thermalize on the prior, 3) global equilibration, we mitigate common pitfalls like mode collapse, variance explosion, mode forgetting, and buffer overfitting. 

With trained models and generated crystal distributions in hand, we undertook distributional analysis based on crystal structural similarity, using the earth mover's distances on atomwise radial distribution functions to define a local probability density, identify the most probable states on the landscape, and draw basins about said maxima.
This allowed us to extract a rich set of observables which are not available in standard CSP searches, including intra-basin statistics and explicit inter-basin probabilities. 
We identified and characterized previously known polymorphic structures, and proposed new crystal structures that are thermodynamically preferred under UMA, according to our model.

As an energy-based model, the distribution sampled by MXtalGFlow is determined by the choice of energy function used in training.
This affords great flexibility to explore the rich interplay between different energy functions and their resulting crystal distributions, an exciting possibility in the era of fast and accurate machine learned potentials.
We exemplified this in our comparison of the distribution of crystals under a Lennard-Jones energy vs the Universal Model for Atoms, where we observed differing degrees of overlap between different molecules based on the differences in their lattice energetics. 

Our work stands in contrast to existing molecular CSP models in several ways, stemming from the observation that molecular crystal polymorphism is driven by inherently statistical quantities, and therefore it should be analyzed on a distributional level.
This idea led to our canonical parameterization scheme and our selection of the energy-based GFlowNet, to model physical distributions of crystals, rather than databases with uncontrolled biases.
These properties give us the ability to rigorously quantify sample diversity, mode coverage, and compute useful thermodynamic observables.
Where prior works use the lattice energy for fine-tuning~\cite{Subramanian2026} or post-sampling crystal optimization, in MXtalGFlow, the energy intrinsically \textit{defines} the distribution of crystals.

Likewise, compared to classical CSP methods, we are able to report physically meaningful distributions of crystals, rather than ranked lists of disconnected structures.
Our thermodynamic observables are then derived directly from the sampled distribution, rather than computed on a subset of promising structures in a separate analysis~\cite{butler2023reducing,hoja2017first}.
Further, once trained, sampling from our model has negligible computational cost, allowing theoretically very high sampling density.

MXtalGFlow is not intended primarily for crystal search but for the sampling and analysis of distributions of crystals.
The clear next step is conditional generalization to a wider range of molecules and space groups with a single model.
Combined with the addition of flexible intramolecular degrees of freedom, such a model could serve as a verifiably diverse and extremely efficient lead generator and probability density estimator across a very broad range of molecular materials.
Further effort should also be spent on improving terminal convergence to improve the quantitative reliability of our thermodynamic predictions.

\section{\label{sec:Methods}Methods}
\subsection{Parameterization}

As a first effort, we here train models on rigid molecules in single space groups.
Flexible intramolecular degrees of freedom can be appended as extra dimensions of the crystal parameterization, given an efficient parameterization and conformer instantiation algorithm, at the cost of increasing problem dimensionality.
The other primary physical constraint of our framework is the assumption of fixed crystal symmetries, due to explicit inclusion only of the asymmetric unit in the parameterization.

As has been explored in the recent inorganic generative CSP literature, space group selectiveness comes with advantages and disadvantages~\cite{Puny2025,Chang2025,AI4Science2023}:  
All-atom diffusion in an unstructured unit cell (implicitly, in the P1 space group) is flexible, yet vulnerable to bias in the training data or the model architecture, which may reduce the probability of sampling certain crystal symmetries.
Further, symmetric crystals (space groups $\in(2,230)$) occupy zero measure in the P1 state space.
Generating symmetric structures in a P1 setting therefore requires post-processing of the sampled crystals, to identify the probable crystal symmetry and snap structures into alignment.
Space group conditioning loses the flexibility and generality of P1 generation but allows users to explicitly target symmetries of interest.

A question arises on how one should interpret a distribution of crystals constrained to remain within a single space group.
High-energy rearrangements (solid-solid phase transitions or re-nucleation events) with barriers well above $k_{\rm B}T$ do not generally follow symmetry-preserving trajectories~\cite{rietveld2023solid,buerger1951phase}.
As such, the space-group-constrained view is not appropriate for such high barrier processes.
However, low-energy, intra-basin fluctuations should be well described in a fixed crystal symmetry. 

Assuming a rigid molecular conformer and a single molecule per asymmetric unit ($Z'=1$), one can completely define a molecular crystal via the space group symmetries and twelve continuous dimensions: the box vector lengths $(a,b,c)$ and interior angles $(\alpha,\beta,\gamma)$, the fractional coordinates of the molecule centroid in the unit cell basis, $(u,v,w)$, and a rotation vector, $(x,y,z)$, defining the rotation of the molecule's canonicalized principal inertial vectors, with respect to the cartesian axes.
In some space groups, the chirality of the molecule in the asymmetric unit is also a free parameter, since, under our parameterization, physical lattices can be expressed with a left or right-handed molecule in the asymmetric unit only in some space groups.
Whether it is possible to express a given lattice with either asymmetric unit chirality comes down to the structure of the normalizers for each space group~\cite{hahn1983international}, with all the Sohncke groups being expressible with either chirality in the asymmetric unit, but many non-Sohncke groups, including $P2_1/c$ not admitting the same.
In such cases, we would train models for both chiralities for a given molecule, and weight them at analysis time by their learned partition functions.

The crystal parameters $\mathcal{C}$ completely describe a rigid molecular crystal in a single space group but do not form a suitable space for sample generation or thermodynamic analysis.
This can be easily seen since, for example, cell lengths and angles have very different numerical ranges and distributions.
A good latent space should ideally be bounded to a given range, cover the search space uniquely, and its euclidean distances should be locally smooth with respect to physical distances.
It should cover the full search space for a given molecule and space group and transfer easily between molecules and space groups.

The core concept of our transform is to reduce crystal parameters to the asymmetric unit.
As all other molecules in the unit cell and crystal lattice are mere symmetry images defined by the space group symmetry; the asymmetric unit is the only free object, assuming a fixed space group.
Accordingly, the cell lengths $(a,b,c)$ and unit cell fractional coordinates $(u,v,w)$ are renormalized into the asymmetric unit via standard conventions~\cite{hahn1983international}.
First, the box vector lengths are divided by the molecule diameter (as defined by twice the distance from the heavy atom centroid to the most distant heavy atom), to generate a molecule-normed length scale.
The normed length is then rescaled from the length of the unit cell to the length of the asymmetric unit, 
\begin{equation}
    a_{latent}=\frac{a\cdot L_{a,au}(SG)}{2\cdot r_{mol}},
\end{equation}
for $r_{mol}$ the molecule radius and $L_{a,au}(SG)$ the length of the asymmetric unit $a$ vector in the fractional basis of the unit cell, for the given space group.
This further normalizes the box vectors between different space groups, which have widely varying symmetries, and therefore contain different numbers of asymmetric units (molecules) in a single unit cell. 

\begin{figure}
\centering
\includegraphics[width=0.5\textwidth]{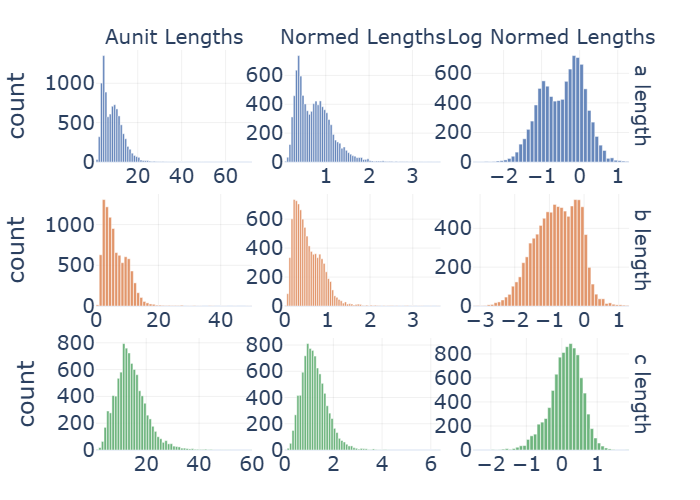}
\caption{\label{fig:csd_lengths} Distributions of asymmetric unit lengths, normed asymmetric unit lengths, and log normed asymmetric unit lengths for 9.4k randomly selected $Z'=1$ structures from the CSD.}
\end{figure}

The distributions of normed asymmetric unit lengths in the Cambridge Structural Database are empirically roughly log-normally distributed, mostly within the range $(0.1, 3)$~\cite{groom2016cambridge}, as we illustrate in Figure~\ref{fig:csd_lengths}.


As we are primarily concerned with the placement of a single molecule in a canonical asymmetric unit, from which the complete unit cell follows by simple application of the space group symmetries, the fractional coordinates of the molecule in the unit cell are likewise renormalized to the asymmetric unit,
\begin{equation}
    u_{au}=\frac{u}{L_{a,au}(SG)}.
\end{equation}
In space group 1, $P1$, the asymmetric unit and unit cell are identical, $L_{abc, au}(1)=(1,1,1)$, and there is no change.
In space group 2, $P\bar{1}$, the asymmetric unit is only half the unit cell length in the $a$ direction, $L_{abc,au}(2)=(0.5, 1, 1)$, so a point at $(0.25, 0.5, 0.5)$ in the unit cell fractional basis would be at $(0.5, 0.5, 0.5)$ in the fractional basis of the asymmetric unit.
Analogously, cell lengths $(a, b, c)$ of the unit cell correspond to asymmetric unit lengths $(\frac{a}{2},b,c)$.

We do not have a na\"ive or physically motivated prior distribution over box parameters or the unit cell metric tensor, so we define such a base distribution using CSD statistics, which cover a very broad range of cell geometries.
We take the logarithm of the normed asymmetric unit lengths, and take this roughly Gaussian object as our base distribution over the cell lengths.
For the rigid body rotation vector $(x,y,z)$, we convert from cartesian to spherical coordinates $(\theta,\phi,r)$, because the periodicity of the rotation vector is more naturally enforceable in this coordinate system.
For representational uniqueness, we restrict the rotation vector to the upper hemisphere, $z\geq 0$, or $\theta\in[0,\frac{\pi}{2}]$. 
The final step in the latent transform is conversion of each parameter to a uniform range, $\mathcal{L}\in[-1,1]$ via linear shifts and rescaling based on their maximum ranges (detailed in SM).
The interior cell angles $(\alpha,\beta,\gamma)$ are left untransformed, except for this rescaling.

Our parameterization is unique in the asymmetric unit degrees of freedom, though for various combinations of molecule and space group symmetries (expressed in the euclidean and affine normalizers), duplicate physical lattices can be realized for different asymmetric unit parameterizations.
As we work in general Wyckoff positions, the action of the normalizer is generically uniform across crystal configurations, so the resulting multiplicities should not, in expectation, distort the sampled populations.

As formulated, this parameterization is not unique in the unit cell box degrees of freedom.
For example, a monoclinic unit cell may be identically represented by an infinite set of box vectors related by a shearing transformation. 
While it is possible to enforce some uniqueness or `standardization' rules intrinsically via the latent transform, this comes at the cost of complexity and generality of the latent basis between crystal systems and, in our experience, all such rules cannot be enforced simultaneously.
Rather than building box uniqueness into the parameterization directly, we define a `standardized' subspace according to common definitions~\cite{kvrivy1976unified,togo2024spglib} and penalize generated samples with large positive energies for lying outside of this subspace.

\subsection{\label{sec:GFN_methods} GFlowNet Model}

Our generative model is a continuous generative flow network (GFlowNet or GFN), structured as a diffusion sampler~\cite{sendera2024improved} formulated as a stochastic differential equation (SDE) of the form
\begin{equation}
{\rm d}{\bm x}_t = \bm{u}_{\theta,F}({\bm x}_t,t){\rm d}t + \bm{g}_{\theta,F}({\bm x}_t,t){\rm d}{\bm w}_t,
\label{eq:GFN}
\end{equation}
where ${\bm x}_t$ is the state of the system at time $t$, ${\bm u}_{\theta,F}({\bm x}_t,t)$ and $g_{\theta,F}({\bm x}_t,t)$ are the drift and diffusion terms, which are trainable functions, approximated by neural networks.
We selected this model due to its flexibility, theoretical properties, and proven capabilities on challenging high-dimensional problems.
GFNs of this type have well defined convergence criteria, admit data or energy-based training, and have few formal limits on the type of diffusion rollout or the architecture of the policy models.
This GFN in Eq.\ (\ref{eq:GFN}) can be written as a discrete time neural SDE using the Euler-Maruyama algorithm, which prescribes the evolution
\begin{equation}
    \bm{x}_{t+\Delta t}=\bm{x}_t+\bm{u}_{\theta,F}(\bm{x}_t, t)\Delta t+\bm{g}_{\theta,F}(\bm{x}_t, t)\odot\bm{z}_t\sqrt{\Delta t},
\end{equation}
where $\Delta t$ is the integration time step, 
$t\in [0,1]$ is the time value, and $\bm{z}_t$ is a vector of standard normal random variables sampled at each discrete time point $t$.
The drift and diffusion terms comprise a forward sampling policy, $p_F$, which generates trajectories $\tau=\{\bm{x}_0,\bm{x}_{\Delta t},{\bm x}_{2\Delta t},...,\bm{x}_1\}$ starting from a constant initial or source state, ${\bm x}_0 = \bm{\delta}_0$, and terminating at distributed samples $\bm{x}_1$.
For training purposes, GFlowNets also employ an auxiliary policy model, the `backwards' policy $p_B$, which samples trajectories along a Brownian bridge from pre-sampled terminal states back to the source state. 

Log probabilities over trajectories are taken sums over discrete steps, 
\begin{equation}
    \log{p(\tau)}=\sum_{n=0}^{T-1}\log{p(\bm{x}_{(n+1)\Delta t}|\bm{x}_{n\Delta t}}),
\end{equation}
with $T$ the number of integration steps such that $T\Delta t = 1$.
We employ two loss functions for different phases of training.
The maximum likelihood estimation (MLE) loss~\cite{zhang2022unifying} is used to train the forward policy to match the distribution of a target dataset, balancing forward and backward policy probabilities on trajectories generated by $p_B$, from terminal states $\bm{x}_1$ sampled from a dataset.
\begin{equation}
    L_{MLE}=\langle\log{p_B(\tau|\bm{x}_T})-\log{p_F(\tau)}\rangle_{\tau\sim p_B(\bm{x}_1)}.
\end{equation}

The second and more characteristic GFlowNet loss is the trajectory balance loss,
\begin{equation}\label{eq:tb}
    L_{TB}=\Big\langle \log{p_F(\tau)} - \log{p_B(\tau)} + \log{Z_\theta} - \log{R(\bm{x}_1)} \Big\rangle^2_{\tau\sim p_F},
\end{equation}
with trajectories sampled under the forward policy, $R(\bm{x}_1)$ a reward associated with the terminal state, and the log partition function, $\log{Z_\theta}$ learned as a self-consistent scalar. 

The TB loss can also be trained in a backward mode, analogous to MLE.
This is useful to equilibrate the model on a given data distribution, and prevent it from forgetting high reward states.
These losses afford GFlowNets the capability to learn a distribution from data, via MLE loss, and also to sharpen and equilibrate a distribution defined by a reward or energy function, via forward and backward TB losses. 
We stage these losses in a robust training workflow, detailed in Section~\ref{sec:protocol}.

\subsection{\label{sec:Prior}Prior Distribution}
In our training workflow, the search space for the GFN model to explore is defined by a `prior' distribution, which is assumed to cover all low-energy packing modes of the crystal.
We synthesize prior datasets for each molecule, space group, and energy function via local optimization of large numbers of randomly initialized crystals on the target energy function, constrained to the standardized box manifold.
The optimization is done directly on the crystal parameters via autograd~\cite{kilgour2025mxtaltoolstoolkitmachinelearning}, yielding, in the limit of large sampling, a reference distribution that defines all areas of nonzero probability density, the support of the target distribution.

Prior datasets were synthesized in $P\bar{1}$ for MIPCAS and $P2_1/c$ for NEHZOR using both Lennard-Jones and UMA potentials.
After filtering and de-duplication, the prior datasets comprised 13.5k and 9.2k crystal samples respectively under LJ and UMA, for MIPCAS, and 26.5k and 5.7k for NEHZOR.
In our workflow, the size of the prior is an intrinsic feature of the crystal landscape, suggesting in this case that the LJ landscapes have significantly more diverse low-energy structures than the UMA landscapes.
This is perhaps unsurprising, as the number of states that are densely packed (LJ minima) likely outnumber the states that are dense, sterically reasonable, and include one or more directional bonds.

This process usually results in significant duplication and sample redundancy. 
We process the raw optimization outputs into a usable prior dataset via the following steps:
1) Identify and remove optimization artifacts, including high-energy states or crystals outside of the standardized manifold.
2) Calibrate the response of low-energy basins to isotropic latent noise.
This is done to define a characteristic length range, $d_{char}$ over which energies increase by a given multiple of $k_BT$.
We empirically observe that physical distances and energies are highly log-linearly correlated with latent noise magnitude over short distances, and so we can easily fit the average noise range over which the energy of a set of low-energy basins will increase by between $0.05k_BT$ and $6k_BT$, $(d_{low},d_{char})$.
3) De-duplication is then done by greedy bottom-up selection.
We take the lowest energy sample $i$, and add it to the prior.
We then take the next lowest energy sample which is at least $d_{char}$ away from sample $i$, add it to the dataset, and so on.
This procedure thins out regions of extreme over-sampling, without erasing real distinct modes of the distribution.

\subsection{\label{sec:Reward}Reward Function}
Energy-based models are trained on a potential energy function such as an empirical force field, machine-learned interatomic potential, score model, or semi-empirical quantum chemical method.
During forward TB training, samples are drawn, crystals constructed, and energies calculated all within the forward pass of the model.
As such, energy evaluation is a key cost in model training.
For benchmarking, contrast, and comparison, we train models on 1) a Lennard-Jones (LJ) type potential with softened repulsion (details in SM), and 2) the Universal Model for Atoms (UMA)~\cite{wood2025family}.
Specifically, we use the UMA `esen-s' checkpoint which has recently been fine-tuned on density functional calculations of 25 million organic molecular crystals~\cite{gharakhanyan2025open}.

Lennard-Jones (LJ) potentials, despite their simplicity, mostly capture the top two factors driving molecular crystal packing, 1) dense packing and 2) minimal intermolecular steric overlap. 
As such, for many molecules, an LJ potential provides at least a plausible starting point for crystal search and refinement. 
Of course, LJ potentials do not describe many important physical phenomena such as hydrogen bonds and all the nuances of electrostatics and dispersion. 
State-of-the-art machine learned potentials like UMA should at least approximately treat such phenomena. 

The primary energy function is then augmented with auxiliary potentials to stabilize training.
These include penalties for unphysically large or small densities, unit cells outside of our standardized subspace, and samples drawn outside of the latent hypercube $(-1,1)^{12}$ (see the SM for details).
For all physically reasonable samples obeying our standardization conditions, these penalties are vanishingly small or zero.
The Jacobian corrections derived in the SM are also added to the base physical energy.
The sample reward is then computed as $R(\bm{x})=\exp\left(-E(\bm{x})/k_BT\right)$.

\subsection{\label{sec:protocol} Training Protocol}
\begin{figure}[!ht]
\centering
\includegraphics[width=\textwidth]{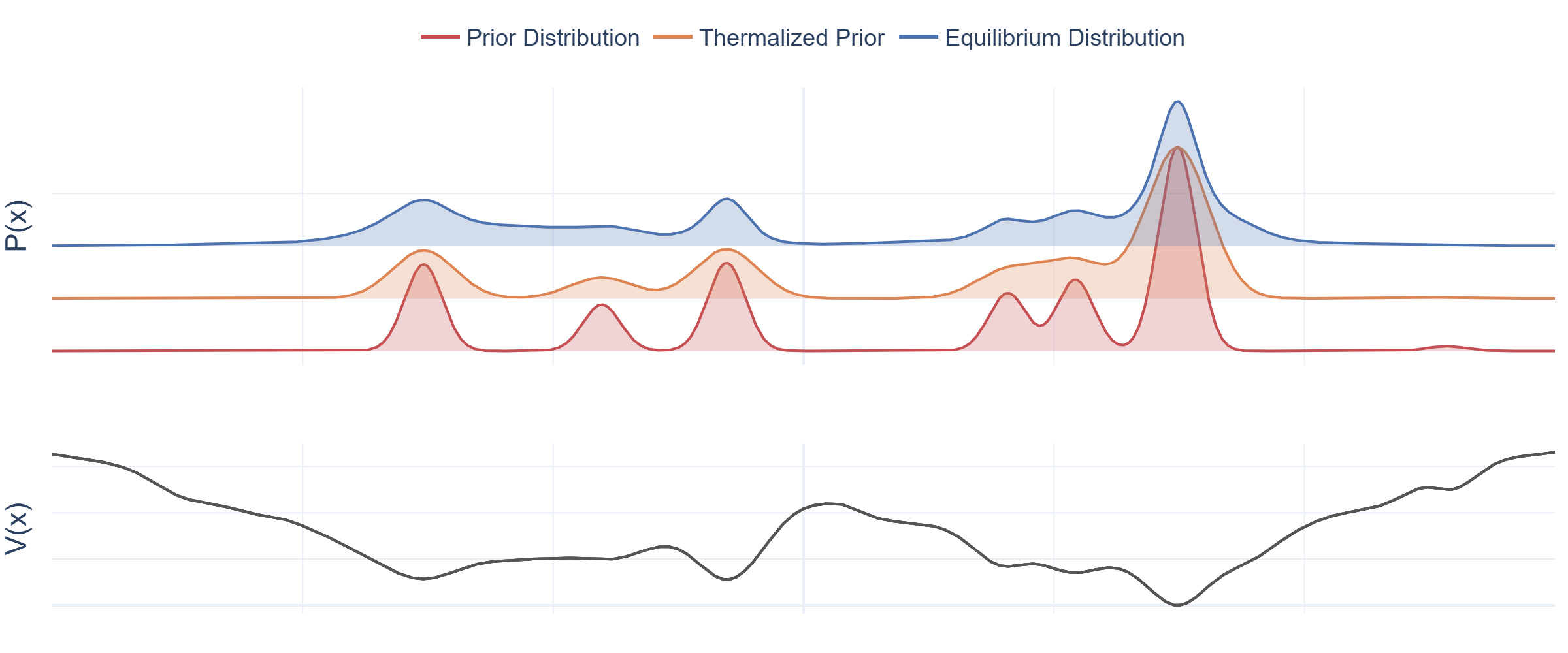}
\caption{\label{fig:pes} Schematic potential energy surface with sample probability distributions for each training phase. Phase 1 uses data-driven training to spread the forward policy to encompass the support of the distribution. Phase 2 equilibrates the backward policy on the prior distribution, and Phase 3 finishes local exploration and equilibration of the forward policy.}
\end{figure}

On-policy TB or VarGrad~\cite{sendera2024improved} losses are generally sufficient to converge GFlowNets on simple systems, particularly on uni-modal or potential landscapes with low barriers compared to $k_BT$.
The high-dimensional multimodality of molecular crystals, combined with their typically rugged potential landscapes, mean that a more advanced training protocol is required in our case.
Our training protocol, diagrammed in Figure~\ref{fig:pes}, satisfies the theoretical requirements for GFlowNet convergence: non-vanishing support on the target distribution and self-consistent convergence of the TB loss.
This is achieved while avoiding common pathologies such as mode collapse, variance explosion, and buffer overfitting.
This systematic approach enables stable and reliable GFlowNet convergence on challenging landscapes.

The first training phase is backwards MLE training, minimizing the KL divergence of the terminal distribution under the forward policy to the prior distribution $D_{KL}(p_F\|p_{\{\bm{x}\}})$, up to a convergence threshold, $\epsilon_{MLE}$.
The purpose of this stage is to guarantee initial distributional support, spreading forward policy distribution over the full range of the prior.
We also warm start learning the partition function on the empirical distribution of the prior.
This is done by computing and back-propagating the TB loss, but only allowing the TB gradients to affect $\log{Z_\theta}$.
This trains $\log{Z_\theta}$ to be the self-consistent normalization factor for the \textbf{current state} of the forward and backward policy models, evaluated on backward trajectories.
In contrast to TB training, MLE training nearly always converges quickly and easily, and is remarkably robust to system and hyperparameter variation, including in model size, energy function, SDE discretization, and parameterization. 

In Phase 2 we introduce sample rewards and smoothly transition to backwards trajectory balance training.
Convergence of Phase 2 corresponds to the equilibration of the policy and $\log{Z_\theta}$, over the states in the prior dataset.
The pathologies of forward TB training (mode collapse, variance explosion) are avoided in backwards-only mode, though with large enough policy models, overfitting to the prior dataset is possible.
Therefore, we train not directly on the prior but on noised and re-evaluated samples from the prior.
The noise magnitude is empirically fit such that the noised samples are mostly within the energetic basin as the prior sample from which they derive.
The `noised buffer' is periodically refreshed to ensure the backwards training has a moving target and so is difficult to overfit (implementation details are given in the SM).

Phase 3 mixes forward and backward TB training, enabling on-policy equilibration of the policy models and terminal convergence towards the Boltzmann distribution.
In contrast to Phase 1, gradients from the backwards TB loss are now allowed to affect the policy models, but not $\log{Z_\theta}$.
Per the trajectory balance conditions, $\log{Z_\theta}$ is now only allowed gradients from the on-policy forward TB loss, on which it will ultimately converge.
In this Phase, gradients from the backward TB loss would bias the $\log{Z_\theta}$ estimate away from the correct on-policy value, towards that for the buffer. 
To avoid mode forgetting/collapse, we dynamically manage the ratio of forward:backward steps, informed by the linear fit of the trajectory balance condition
\begin{equation}
\label{eq:tb_parity}
    \log R(\bm{x}_T) + \log p_B(\tau) \quad vs. \quad \log p_F(\tau|\bm{x}_T) + \log Z_\theta,
\end{equation}
separately evaluated on-policy under the forward model, and on the prior dataset.
Forward training is restricted until the slope and intercept of this fit on the prior satisfy pre-set convergence criteria. 
This, combined with the noising of the prior states in backward training, prevents both mode collapse (forgetting) and overfitting.

\section*{Supporting Material}
We attach extensive supporting material, including additional experimental figures, derivations, definitions, and implementation details.

\section*{Acknowledgements}
MK, JR, and MET acknowledge funding from the National Science Foundation, DMR-2118890, and MET from CHE-1955381. This work was supported in part through the NYU IT High Performance Computing resources, services, and staff expertise.
The Flatiron Institute is a division of the Simons Foundation.

\bibliographystyle{ieeetr}
\bibliography{main}

@inproceedings{jain2022biological,
	title = {Biological sequence design with gflownets},
	author = {Jain, Moksh and Bengio, Emmanuel and Hernandez-Garcia, Alex and Rector-Brooks, Jarrid and Dossou, Bonaventure FP and Ekbote, Chanakya Ajit and Fu, Jie and Zhang, Tianyu and Kilgour, Michael and Zhang, Dinghuai and others},
	booktitle = {International Conference on Machine Learning},
	doi = {10.48550/arxiv.2203.04115},
	editor = {Chaudhuri, Kamalika and Jegelka, Stefanie and Le Song and Szepesvári, Csaba and Niu, Gang and Sabato, Sivan},
	journal = {arXiv},
	month = {3},
	organization = {PMLR},
	pages = {9786--9801},
	publisher = {Cornell University},
	url = {https://arxiv.org/pdf/2203.04115},
	volume = {162},
	year = {2022},
}

@article{kilgour2025multi,
	title = {Multi-type point cloud autoencoder: a complete equivariant embedding for molecule conformation and pose},
	author = {Kilgour, Michael and Tuckerman, Mark E and Rogal, Jutta},
	doi = {10.1088/2632-2153/adff35},
	issn = {2632-2153},
	journal = {Machine Learning: Science and Technology},
	month = {8},
	number = {3},
	pages = {035055},
	publisher = {IOP Publishing},
	volume = {6},
	year = {2025},
}

@article{sendera2024improved,
	title = {Improved off-policy training of diffusion samplers},
	author = {Sendera, Marcin and Kim, Minsu and Mittal, Sarthak and Lemos, Pablo and Scimeca, Luca and Rector-Brooks, Jarrid and Adam, Alexandre and Bengio, Yoshua and Malkin, Nikolay},
	booktitle = {Advances in Neural Information Processing Systems 37},
	doi = {10.52202/079017-2575},
	editor = {Globersons, Amir and Mackey, Lester and Belgrave, Danielle and Fan, Angela and Paquet, Ulrich and Tomczak, Jakub M. and Zhang, Cheng},
	journal = {Advances in Neural Information Processing Systems},
	month = {2},
	pages = {81016--81045},
	publisher = {Neural Information Processing Systems Foundation, Inc. (NeurIPS)},
	volume = {37},
	year = {2024},
}

@inproceedings{lahlou2023theory,
	title = {A theory of continuous generative flow networks},
	author = {Lahlou, Salem and Deleu, Tristan and Lemos, Pablo and Zhang, Dinghuai and Volokhova, Alexandra and Hernandez-Garcia, Alex and Ezzine, L{\'e}na N{\'e}hale and Bengio, Yoshua and Malkin, Nikolay},
	booktitle = {International Conference on Machine Learning},
	doi = {10.48550/arxiv.2301.12594},
	editor = {Krause, Andreas and Brunskill, Emma and Cho, KyungHyun and Engelhardt, Barbara and Sabato, Sivan and Scarlett, Jonathan},
	journal = {arXiv},
	month = {1},
	organization = {PMLR},
	pages = {18269--18300},
	publisher = {Cornell University},
	url = {https://arxiv.org/pdf/2301.12594},
	volume = {abs/2301.12594},
	year = {2023},
}

@article{bengio2023gflownet,
	title = {Gflownet foundations},
	author = {Bengio, Yoshua and Lahlou, Salem and Deleu, Tristan and Hu, Edward J and Tiwari, Mo and Bengio, Emmanuel},
	journal = {Journal of Machine Learning Research},
	number = {210},
	pages = {1--55},
	volume = {24},
	year = {2023},
}

@article{zhang2022unifying,
	title = {Unifying generative models with GFlowNets and beyond},
	author = {Zhang, Dinghuai and Chen, Ricky TQ and Malkin, Nikolay and Bengio, Yoshua},
	doi = {10.48550/arxiv.2209.02606},
	journal = {arXiv preprint arXiv:2209.02606},
	month = {9},
	publisher = {Cornell University},
	url = {https://arxiv.org/pdf/2209.02606},
	year = {2022},
}

@article{kvrivy1976unified,
	title = {A unified algorithm for determining the reduced (Niggli) cell},
	author = {K{\v{r}}ivy, I and Gruber, B},
	issn = {0567-7394},
	journal = {Foundations of Crystallography},
	month = {3},
	number = {2},
	pages = {297--298},
	publisher = {International Union of Crystallography},
	volume = {32},
	year = {1976},
}

@article{groom2016cambridge,
	title = {The Cambridge structural database},
	author = {Groom, Colin R and Bruno, Ian J and Lightfoot, Matthew P and Ward, Suzanna C},
	issn = {2052-5192},
	journal = {Acta Crystallographica Section B: Structural Science, Crystal Engineering and Materials},
	month = {4},
	number = {2},
	pages = {171--179},
	publisher = {International Union of Crystallography},
	url = {https://journals.iucr.org/b/issues/2016/02/00/bm5086/bm5086.pdf},
	volume = {72},
	year = {2016},
}

@misc{jin2025oxtalallatomdiffusionmodel,
	title = {OXtal: An All-Atom Diffusion Model for Organic Crystal Structure Prediction},
	author = {Emily Jin and Andrei Cristian Nica and Mikhail Galkin and Jarrid Rector-Brooks and Kin Long Kelvin Lee and Santiago Miret and Frances H. Arnold and Michael Bronstein and Avishek Joey Bose and Alexander Tong and Cheng-Hao Liu},
	archiveprefix = {arXiv},
	doi = {10.48550/arxiv.2512.06987},
	eprint = {2512.06987},
	journal = {arXiv},
	month = {12},
	primaryclass = {cs.LG},
	publisher = {Cornell University},
	url = {https://arxiv.org/abs/2512.06987},
	year = {2025},
}

@article{togo2024spglib,
	title = {Spglib: a software library for crystal symmetry search},
	author = {Togo, Atsushi and Shinohara, Kohei and Tanaka, Isao},
	doi = {10.1080/27660400.2024.2384822},
	issn = {2766-0400},
	journal = {Science and Technology of Advanced Materials: Methods},
	month = {8},
	number = {1},
	pages = {2384822},
	publisher = {Taylor \& Francis},
	url = {https://www.tandfonline.com/doi/pdf/10.1080/27660400.2024.2384822?needAccess=true},
	volume = {4},
	year = {2024},
}

@book{hahn1983international,
	title = {International tables for crystallography},
	author = {Hahn, Theo and Shmueli, Uri and Arthur, JC Wilson},
	publisher = {Reidel Dordrecht},
	volume = {1},
	year = {1983},
}

@article{kilgour2025mxtaltoolstoolkitmachinelearning,
author = {Kilgour, Michael and Tuckerman, Mark E. and Rogal, Jutta},
title = {MXtalTools: A Toolkit for Machine Learning on Molecular Crystals},
journal = {Journal of Chemical Information and Modeling},
volume = {66},
number = {7},
pages = {3402-3408},
year = {2026},
doi = {10.1021/acs.jcim.5c02868},
    note ={PMID: 41875147},
URL = {https://doi.org/10.1021/acs.jcim.5c02868},
eprint = {https://doi.org/10.1021/acs.jcim.5c02868}
}

@article{gharakhanyan2025open,
	title = {Open molecular crystals 2025 (omc25) dataset and models},
	author = {Gharakhanyan, Vahe and Barroso-Luque, Luis and Yang, Yi and Shuaibi, Muhammed and Michel, Kyle and Levine, Daniel S and Dzamba, Misko and Fu, Xiang and Gao, Meng and Liu, Xingyu and others},
	doi = {10.48550/arxiv.2508.02651},
	journal = {arXiv preprint arXiv:2508.02651},
	month = {8},
	publisher = {Cornell University},
	url = {https://arxiv.org/pdf/2508.02651},
	year = {2025},
}

@article{wood2025family,
	title = {UMA: A Family of Universal Models for Atoms},
	author = {Wood, Brandon M and Dzamba, Misko and Fu, Xiang and Gao, Meng and Shuaibi, Muhammed and Barroso-Luque, Luis and Abdelmaqsoud, Kareem and Gharakhanyan, Vahe and Kitchin, John R and Levine, Daniel S and others},
	booktitle = {Advances in Neural Information Processing Systems 38: Annual Conference on Neural Information Processing Systems 2025, NeurIPS 2025, San Diago, CA, USA, December 2-7, 2025 / Mexico City, Mexico, November 30 - December 5, 2025},
	doi = {10.48550/arxiv.2506.23971},
	editor = {Belgrave, Danielle and Zhang, Cheng and Montoya, Laura N. and Lin, Hsuan-Tien and Pascanu, Razvan and Koniusz, Piotr and Ghassemi, Marzyeh and Chen, Nancy and Ruíz, Iván Vladimir Meza and Loaiza-Bonilla, Arturo},
	issn = {2331-8422},
	journal = {arXiv preprint arXiv:2506.23971},
	month = {6},
	publisher = {Cornell University},
	url = {https://arxiv.org/pdf/2506.23971},
	volume = {abs/2506.23971},
	year = {2025},
}

@article{sykes2024has,
	title = {What has scripting ever done for us? The CSD Python application programming interface (API)},
	author = {Sykes, Richard A and Johnson, Natalie T and Kingsbury, Christopher J and Harter, J{\"u}rgen and Maloney, Andrew GP and Sugden, Isaac J and Ward, Suzanna C and Bruno, Ian J and Adcock, Stewart A and Wood, Peter A and others},
	issn = {0021-8898},
	journal = {Applied Crystallography},
	month = {7},
	number = {4},
	pages = {1235--1250},
	publisher = {International Union of Crystallography},
	volume = {57},
	year = {2024},
}

@article{hoja2017first,
	title = {First-principles modeling of molecular crystals: structures and stabilities, temperature and pressure},
	author = {Hoja, Johannes and Reilly, Anthony M and Tkatchenko, Alexandre},
	issn = {1759-0876},
	journal = {Wiley Interdisciplinary Reviews: Computational Molecular Science},
	month = {1},
	number = {1},
	pages = {e1294},
	publisher = {Wiley Online Library},
	volume = {7},
	year = {2017},
}

@article{mcinnes2018umap,
	title = {Umap: Uniform manifold approximation and projection for dimension reduction},
	author = {McInnes, Leland and Healy, John and Melville, James},
	doi = {10.48550/arxiv.1802.03426},
	journal = {arXiv preprint arXiv:1802.03426},
	month = {2},
	publisher = {Cornell University},
	url = {https://arxiv.org/pdf/1802.03426},
	year = {2018},
}

@article{Subramanian2026,
	title = {{PackFlow: Generative Molecular Crystal Structure Prediction via Reinforcement Learning Alignment}},
	author = {Subramanian, Akshay and Pan, Elton and Nam, Juno and Weiler, Maurice and Qu, Shuhui and Park, Cheol Woo and Jaakkola, Tommi S. and Olivetti, Elsa and Gomez-Bombarelli, Rafael},
	archiveprefix = {arXiv},
	arxivid = {2602.20140},
	eprint = {2602.20140},
	journal = {arXiv},
	publisher = {Cornell University},
	url = {http://arxiv.org/abs/2602.20140},
	year = {2026},
}

@article{Zeng,
	title = {{MolCrystalFlow: Molecular Crystal Structure Prediction via Flow Matching}},
	author = {Zeng, Cheng and Sullivan, Harry W and Egg, Thomas and Martirossyan, Maya M and H{\"{o}}llmer, Philipp and Jin, Jirui and Hennig, Richard G and Roitberg, Adrian and Martiniani, Stefano and Tadmor, Ellad B and Liu, Mingjie},
	archiveprefix = {arXiv},
	arxivid = {arXiv:2602.16020v1},
	doi = {10.48550/arxiv.2602.16020},
	eprint = {arXiv:2602.16020v1},
	journal = {Open MIND},
	month = {2},
	pages = {1--20},
	year = {2026},
}

@book{bernstein2020polymorphism,
	title = {Polymorphism in molecular crystals},
	author = {Bernstein, Joel},
	isbn = {978-0199655441},
	month = {5},
	publisher = {Oxford University Press},
	volume = {30},
	year = {2020},
}

@article{metni2026generative,
	title = {Generative models for crystalline materials},
	author = {Metni, Houssam and Ruple, Laura and Walters, Lauren N and Torresi, Luca and Teufel, Jonas and Schopmans, Henrik and {\"O}streicher, Jona and Zhang, Yumeng and Neubert, Marlen and Koide, Yuri and others},
	issn = {0935-9648},
	journal = {Advanced Materials},
	month = {2},
	number = {18},
	pages = {e23620},
	publisher = {Wiley Online Library},
	volume = {38},
	year = {2026},
}

@inproceedings{klipfel2024vector,
author = {Klipfel, Astrid and Fregier, Ya\"{e}l and Sayede, Adlane and Bouraoui, Zied},
title = {Vector field oriented diffusion model for crystal material generation},
year = {2024},
isbn = {978-1-57735-887-9},
publisher = {AAAI Press},
url = {https://doi.org/10.1609/aaai.v38i20.30224},
doi = {10.1609/aaai.v38i20.30224},
booktitle = {Proceedings of the Thirty-Eighth AAAI Conference on Artificial Intelligence and Thirty-Sixth Conference on Innovative Applications of Artificial Intelligence and Fourteenth Symposium on Educational Advances in Artificial Intelligence},
articleno = {2476},
numpages = {9},
series = {AAAI'24/IAAI'24/EAAI'24}
}

@article{Puny2025,
	title = {{Space Group Conditional Flow Matching}},
	author = {Puny, Omri and Lipman, Yaron and Miller, Benjamin Kurt},
    journal = {arXiv},
	archiveprefix = {arXiv},
	arxivid = {2509.23822},
	eprint = {2509.23822},
	url = {http://arxiv.org/abs/2509.23822},
	year = {2025},
}

@article{Chang2025,
	title = {{Space Group Equivariant Crystal Diffusion}},
	author = {Chang, Rees and Pak, Angela and Guerra, Alex and Zhan, Ni and Richardson, Nick and Ertekin, Elif and Adams, Ryan P.},
	archiveprefix = {arXiv},
	arxivid = {2505.10994},
	doi = {10.48550/arxiv.2505.10994},
	eprint = {2505.10994},
	journal = {arXiv},
	month = {5},
	publisher = {Cornell University},
	url = {http://arxiv.org/abs/2505.10994},
	year = {2025},
}

@article{AI4Science2023,
	title = {{Crystal-GFN: sampling crystals with desirable properties and constraints}},
	author = {Hernandez-Garcia, Alex and Duval, Alexandre and Volokhova, Alexandra and Bengio, Yoshua and Sharma, Divya and Carrier, Pierre Luc and Benabed, Yasmine and Koziarski, Micha{\l} and Schmidt, Victor},
	archiveprefix = {arXiv},
	arxivid = {2310.04925},
	eprint = {2310.04925},
	journal = {arXiv},
	month = {10},
	number = {Ml},
	publisher = {Cornell University},
	url = {http://arxiv.org/abs/2310.04925},
	year = {2023},
}

@article{rietveld2023solid,
	title = {Solid-solid phase transitions between crystalline polymorphs of organic materials},
	author = {Rietveld, Ivo B},
	issn = {1381-6128},
	journal = {Current Pharmaceutical Design},
	month = {2},
	number = {6},
	pages = {445--461},
	publisher = {Bentham Science Publishers direct},
	volume = {29},
	year = {2023},
}

@article{buerger1951phase,
	title = {Phase transformations in solids},
	author = {Buerger, Martin J},
	journal = {John Wiley},
	pages = {183},
	year = {1951},
}

@article{tuckerman2023topological,
	title = {Topological Crystal Structure Prediction},
	author = {Tuckerman, Mark and Galanakis, Nikolaos},
	journal = {Research Square},
	month = {9},
	url = {https://www.researchsquare.com/article/rs-3361974/latest.pdf},
	year = {2023},
}

@article{buckingham1938classical,
	title = {The classical equation of state of gaseous helium, neon and argon},
	author = {Buckingham, Richard A},
	journal = {Proceedings of the Royal Society of London. Series A. Mathematical and Physical Sciences},
	month = {10},
	number = {933},
	pages = {264--283},
	publisher = {The Royal Society London},
	volume = {168},
	year = {1938},
}

@book{kaplan2006intermolecular,
	title = {Intermolecular interactions: physical picture, computational methods and model potentials},
	author = {Kaplan, Ilʹi︠a︡ Grigorʹevich and others},
	publisher = {Wiley Online Library},
	volume = {367},
	year = {2006},
}

@article{noe2019boltzmann,
	title = {Boltzmann generators: Sampling equilibrium states of many-body systems with deep learning},
	author = {No{\'e}, Frank and Olsson, Simon and K{\"o}hler, Jonas and Wu, Hao},
	issn = {0036-8075},
	journal = {Science},
	month = {9},
	number = {6457},
	pages = {eaaw1147},
	publisher = {American Association for the Advancement of Science},
	url = {https://science.sciencemag.org/content/sci/365/6457/eaaw1147.full.pdf},
	volume = {365},
	year = {2019},
}

@article{chisholm2005compack,
	title = {COMPACK: a program for identifying crystal structure similarity using distances},
	author = {Chisholm, James Alexander and Motherwell, Sam},
	journal = {Applied Crystallography},
	number = {1},
	pages = {228--231},
	publisher = {International Union of Crystallography},
	volume = {38},
	year = {2005},
}

@article{bengio2021flow,
	title = {Flow network based generative models for non-iterative diverse candidate generation},
	author = {Bengio, Emmanuel and Jain, Moksh and Korablyov, Maksym and Precup, Doina and Bengio, Yoshua},
	booktitle = {Advances in Neural Information Processing Systems 34: Annual Conference on Neural Information Processing Systems 2021, NeurIPS 2021, December 6-14, 2021, virtual},
	doi = {10.48550/arxiv.2106.04399},
	editor = {Ranzato, Marc'Aurelio and Beygelzimer, Alina and Dauphin, Yann N. and Liang, Percy and Vaughan, Jennifer Wortman},
	journal = {Advances in neural information processing systems},
	month = {6},
	pages = {27381--27394},
	publisher = {Cornell University},
	url = {https://arxiv.org/pdf/2106.04399},
	volume = {34},
	year = {2021},
}

@article{butler2023reducing,
	title = {Reducing overprediction of molecular crystal structures via threshold clustering},
	author = {Butler, Patrick WV and Day, Graeme M},
	doi = {10.1073/pnas.2300516120},
	issn = {0027-8424},
	journal = {Proceedings of the National Academy of Sciences},
	month = {5},
	number = {23},
	pages = {e2300516120},
	publisher = {National Academy of Sciences},
	url = {https://pmc.ncbi.nlm.nih.gov/articles/PMC10266058/pdf/pnas.202300516.pdf},
	volume = {120},
	year = {2023},
}

@article{vahediahmar2026orgflow,
	title = {OrgFlow: Generative Modeling of Organic Crystal Structures from Molecular Graphs},
	author = {Vahediahmar, Mohammadmahdi and McDonald, Matthew A and Liu, Feng},
	journal = {arXiv preprint arXiv:2602.20195},
	month = {2},
	publisher = {Cornell University},
	url = {https://arxiv.org/pdf/2602.20195},
	year = {2026},
}

@article{yershova2010generating,
	title = {Generating uniform incremental grids on SO (3) using the Hopf fibration},
	author = {Yershova, Anna and Jain, Swati and Lavalle, Steven M and Mitchell, Julie C},
	journal = {The International journal of robotics research},
	number = {7},
	pages = {801--812},
	publisher = {SAGE Publications Sage UK: London, England},
	volume = {29},
	year = {2010},
}

@article{Kwon2026,
	title = {{Stable-GFlowNet: Toward Diverse and Robust LLM Red-Teaming via Contrastive Trajectory Balance}},
	author = {Kwon, Minchan and Baek, Sunghyun and Kim, Minseo and Yu, Jaemyung and Han, Dongyoon and Kim, Junmo},
	archiveprefix = {arXiv},
	arxivid = {2605.00553},
	eprint = {2605.00553},
	journal = {arXiv},
	month = {5},
	publisher = {Cornell University},
	url = {http://arxiv.org/abs/2605.00553},
	year = {2026},
}

@article{Morozov2026,
	title = {{Learning Shortest Paths with Generative Flow Networks}},
	author = {Morozov, Nikita and Maksimov, Ian and Tiapkin, Daniil and Samsonov, Sergey},
	archiveprefix = {arXiv},
	arxivid = {2603.01786},
	eprint = {2603.01786},
	journal = {arXiv},
	month = {3},
	publisher = {Cornell University},
	url = {http://arxiv.org/abs/2603.01786},
	year = {2026},
}

@article{Niu2026,
	title = {{Evaluating GFlowNet from partial episodes for stable and flexible policy-based training}},
	author = {Niu, Puhua and Wu, Shili and Qian, Xiaoning},
	archiveprefix = {arXiv},
	arxivid = {2603.01047},
	doi = {10.48550/arxiv.2603.01047},
	eprint = {2603.01047},
	issn = {2331-8422},
	journal = {arXiv},
	month = {3},
	publisher = {Cornell University},
	url = {http://arxiv.org/abs/2603.01047},
	volume = {abs/2603.01047},
	year = {2026},
}

@article{Yu2026,
	title = {{Partial GFlowNet: Accelerating Convergence in Large State Spaces via Strategic Partitioning}},
	author = {Yu, Xuan and Wang, Xu and Zhu, Rui and Zhang, Yudong and Wang, Yang},
	archiveprefix = {arXiv},
	arxivid = {2602.11498},
	doi = {10.48550/arxiv.2602.11498},
	eprint = {2602.11498},
	journal = {Open MIND},
	month = {2},
	url = {http://arxiv.org/abs/2602.11498},
	year = {2026},
}

@article{Chen2026,
	title = {{Controlling Exploration-Exploitation in GFlowNets via Markov Chain Perspectives}},
	author = {Chen, Lin and Drapeau, Samuel and Shao, Fanghao and Zhu, Xuekai and Xue, Bo and Song, Yunchong and Lauri{\`{e}}re, Mathieu and Lin, Zhouhan},
	archiveprefix = {arXiv},
	arxivid = {2602.01749},
	eprint = {2602.01749},
	issn = {2331-8422},
	journal = {Open MIND},
	month = {2},
	url = {http://arxiv.org/abs/2602.01749},
	volume = {abs/2602.01749},
	year = {2026},
}

@article{Malek2025,
	title = {{Loss-Guided Auxiliary Agents for Overcoming Mode Collapse in GFlowNets}},
	author = {Malek, Idriss and Sharma, Abhijit and Lahlou, Salem},
	archiveprefix = {arXiv},
	arxivid = {2505.15251},
	doi = {10.1609/aaai.v40i29.39613},
	eprint = {2505.15251},
	issn = {23743468},
	journal = {arXiv},
	month = {5},
	publisher = {Cornell University},
	url = {http://arxiv.org/abs/2505.15251},
	year = {2025},
}

@article{Rector-Brooks2023,
	title = {{Thompson sampling for improved exploration in GFlowNets}},
	author = {Rector-Brooks, Jarrid and Madan, Kanika and Jain, Moksh and Korablyov, Maksym and Liu, Cheng-Hao and Chandar, Sarath and Malkin, Nikolay and Bengio, Yoshua},
	archiveprefix = {arXiv},
	arxivid = {2306.17693},
	doi = {10.48550/arxiv.2306.17693},
	eprint = {2306.17693},
	journal = {arXiv},
	month = {6},
	publisher = {Cornell University},
	url = {http://arxiv.org/abs/2306.17693},
	year = {2023},
}

@article{lo2026fast,
	title = {Fast Organic Crystal Structure Prediction with Unit Cell Flow Matching},
	author = {Lo, Alston and Mucko, Luka and Cheng, Austin H and Cai, Andy and Price, Alastair JA and Matusik, Wojciech and Aspuru-Guzik, Al{\'a}n},
	doi = {10.48550/arxiv.2606.03199},
	journal = {arXiv preprint arXiv:2606.03199},
	month = {6},
	publisher = {Cornell University},
	year = {2026},
}

@article{yang2022global,
	title = {Global analysis of the energy landscapes of molecular crystal structures by applying the threshold algorithm},
	author = {Yang, Shiyue and Day, Graeme M},
	doi = {10.1038/s42004-022-00705-4},
	issn = {2399-3669},
	journal = {Communications Chemistry},
	month = {7},
	number = {1},
	pages = {86},
	publisher = {Nature Publishing Group UK London},
	url = {https://www.nature.com/articles/s42004-022-00705-4.pdf},
	volume = {5},
	year = {2022},
}

@article{yang2021exploration,
	title = {Exploration and optimization in crystal structure prediction: Combining basin hopping with quasi-random sampling},
	author = {Yang, Shiyue and Day, Graeme M},
	issn = {1549-9618},
	journal = {Journal of Chemical Theory and Computation},
	month = {2},
	number = {3},
	pages = {1988--1999},
	publisher = {ACS Publications},
	volume = {17},
	year = {2021},
}

\renewcommand{\thefigure}{S\arabic{figure}}
\renewcommand{\thetable}{S\arabic{table}}
\renewcommand{\theequation}{S\arabic{equation}}

\renewcommand{\thesection}{S\arabic{section}}
\setcounter{section}{0}  
\setcounter{figure}{0}  
\setcounter{table}{0}  
\setcounter{equation}{0}  

\onecolumn
\section{\label{sec:architecture} Model Details}
Our GFlowNet model follows closely the implementation in~\cite{sendera2024improved}.
A forward SDE step is integrated from a constant initial state, $\bm{x}_0=\bm{\delta}_0$, according to,
\begin{equation}
        \bm{x}_{t+\Delta t}=\bm{x}_t+\bm{u}_F \Delta t+\bm{g}_F\odot\bm{z}\sqrt{\Delta t},
\end{equation}
with $\bm{u}_F$, $\bm{g}_F$ the 12 dimensional drift and diffusion terms.
The policy terms are outputs from the forward policy model,
\begin{equation}
    \bm{u}_F, \bm{d}_F=P_{F,\theta}(\bm{x}_{emb}, \bm{t}_{emb}),
\end{equation}
where $\bm{d}_F$ is the unscaled diffusion coefficient, $\bm{x}_{emb}$ is the output of the state embedding model, $\bm{x}_{emb}=S_\theta(\bm{x}_t)$, and $\bm{t}_{emb}$ is the output of the time embedding model $\bm{t}_{emb}=T_\theta(t)$.
To improve model stability, the maximum available range of the diffusion coefficient is gated according to,
\begin{equation}
    \log{\bm{g}_F^2}=\log{\sigma^2_0}+\tanh\left(\frac{\bm{d}_F}{s_{max}}\right) s_{max},
\end{equation}
with $\sigma^2_0$ a constant baseline policy variance, and $s_{max}$ the maximum range about the baseline the model is allowed to explore.
Periodic angular degrees of freedom, in this work the azimuthal angle $\phi$ and rotation vector length $r$ are each embedded as $e_{r,1}, e_{r,2}=\sin r, \cos r,\ e_{\phi,1}, e_{\phi,2}=\sin \phi, \cos \phi$ before passing to the state embedding model.

We parameterize the GFlowNet policy using geometry-aware stochastic transition kernels. 
Non-periodic continuous variables are propagated via Gaussian transitions in Euclidean space, while periodic variables are modeled using wrapped kernels, ensuring continuity and normalization under periodic boundary conditions. 
For non-periodic variables, the forward transition log-density is given for dimension $i$ by
\begin{equation}
\begin{split}
	\log p (x_{t+\Delta t,i} | x_{t,i}) &= \log \mathcal{N} \left( x_{t+\Delta t,i} | x_{t,i}+u_{i}\Delta t,\ g_{i}^2\Delta t\right),\\
    &= -\frac{1}{2}\left(z_i^2 + \log(g_i^2\Delta t)+\log(2\pi)\right).
\end{split}
\end{equation}
with $\bm{z}$, the noise in the propagation step, recomputed for gradient flow using reparameterization trick,
\begin{equation}
    z_i=\frac{x_{t+\Delta t,i}-x_{t,i} - u_i\Delta t}{g_i\sqrt{\Delta t}}.
\end{equation}
For periodic variables, we approximate the wrapped transition kernel using a single Gaussian evaluated on wrapped angular differences,
\begin{equation}
	\log p (x_{t+\Delta t,i} | x_{t,i}) = -\frac{1}{2} \left[ \left(\frac{\text{wrap}_{\pi}(x_{t+\Delta t,i} - x_{t,i} - u_{i}\Delta t)}{g_{i}\sqrt{\Delta t}}\right)^2 + \log (g_{i}^2\Delta t) + \log 2\pi \right],
\end{equation}
which avoids discontinuities at angular boundaries. 
This approximation is valid in the small step-size regime where probability mass from multiple windings is negligible.

Following the standard method given in~\cite{sendera2024improved}, we do not by default allow gradients to flow through the diffusion rollout directly, rather, gradients are computed directly on the policy model outputs.
We find empirically however, that convergence on the MLE training task is greatly helped by through-trajectory gradients, and so we enable this in our off-policy training.

Backwards SDE steps are analogous to forward, with a few adjustments.
Trajectories are sampled from a given terminal state, $\bm{x}_T$, stepwise according to
\begin{equation}
        \bm{x}_{t}=\bm{x}_{t+\Delta t}+\bm{u}_B\Delta t + \bm{g}_B\odot\bm{z}\sqrt{\Delta t}.
\end{equation}
The backward policy model is a Brownian bridge, a stochastic process with fixed endpoints.
The drift points deterministically towards the source node (usually implicit at 0),
\begin{equation}
    \bm{u}_B=- \frac{1}{t}\bm{x}_{t+\Delta t}\odot\bm{c}_{\theta,u}(\bm{x}_{emb},\bm{t}_{emb}),
\end{equation}
with $\bm{c}_{\theta,u}$ a learned linear correction, output from the backward policy model analogously to the forward drift, and $\bm{x}_{emb}, \bm{t}_{emb}$ embedded at step $x_{t+\Delta t}, t+\Delta t$.
The backwards policy variance also has a learned part and a deterministic part which converges to zero for $t=\Delta t$,
\begin{equation}
    \bm{g}_B^2=\sigma_0^2 \bm{c}_{\theta,\sigma}(\bm{x}_{emb},\bm{t}_{emb})\frac{(t-\Delta t)}{t},
\end{equation}
with $\bm{c}_{\theta,\sigma}$ a learned correction from the backward policy model.

All the learning models are standard multilayer perceptrons with residual connections, as implemented in the MXtalTools package~\cite{kilgour2025mxtaltoolstoolkitmachinelearning}, and the partition function estimate is a learnable scalar.
Separate Adam optimizers are used for forward and backward training steps, as well as the partition function.
The time and state embedding models are shared between forward and backward SDE integration.

Trajectory balance losses minimize the absolute log residual on a given trajectory,
\begin{equation}
    r_\Delta=|\log{p_F(\tau)} - \log{p_B(\tau|\bm{x}_T)} + \log{Z_\theta} - \log{R(\bm{x}_T)}|.
\end{equation}
Traditionally this is optimized against a quadratic penalty, $L_{TB}=r_\Delta^2$, but for systems with wide ranges of rewards, this can lead to training instability.
We therefore use a smooth L1 type (or Huber type) loss with a very generous cutoff, $\beta=10$, which provides a wide quadratic-like basin, while avoiding loss explosions when there are large residuals,
\begin{equation}
L_{TB} =
\begin{cases}
\frac{r_\Delta^2}{2\beta}, & |r_\Delta| < \beta, \\
|r_\Delta|-\frac{1}{2}\beta, & |r_\Delta| > \beta.
\end{cases}
\end{equation}

The model and SDE hyperparameters are as follows:
\begin{itemize}
    \item Forward and backward policy models, as well as the state embedding model are 4 layer MLPs with 1024 hidden dimensions, skip connections, and layer normalization.
    \item Diffusion rollout is done on a static and evenly spaced 100-step grid
    \item Base policy variance is 0.05. Forward and backward policies can learn corrections in the range of $\pm6$ log units. The backwards policy can learn corrections to the drift with a maximum magnitude of 0.2 times the Brownian bridge drift magnitude.
    \item The final evaluation model is trained as an exponential moving average with respect to the training model with a decay value of 0.95. 
    \item Forward and backward policy learning rates are initialized at $4e^{-6}$, ramped exponentially to a maximum of $5e^{-4}$ over 2000 steps or until loss instability, then decay exponentially to $1e^{-5}$ over 100k steps. The $\log{Z_\theta}$ has a separate optimizer and learning rate fixed at 1.0.
    \item Batch sizes are initialized at 50 and grow up to a max of 3000 or GPU VRAM saturation. 
    \item GFN models were trained on cluster A100 GPUs for 4 days. 
\end{itemize}

\section{Prior Noising}
We mitigate overfitting of the prior by noising and re-scoring the samples used during backward training via $\bm{x}_{noised}=\bm{v}+c_n\cdot\bm{\hat{n}}$, where $c_n$ is log-uniform noise on the range $(d_{low},d_{char})$, and $\hat{n}$ is a random unit vector.
The noise range is calibrated such that noised samples are mostly within the same energetic basins as the original samples, giving the model a picture of the energy landscape near each point, not just the local minimum.

A large set of noisy states (200k in our experiments) are generated before training begins.
Even with a large number of noised states, the model may still overfit a static target, so the noised buffer refreshes dynamically over time, with older and lower-loss samples exiting preferentially.

New samples are selected based on the states in the prior that are under-weighted by the model.
At each evaluation step, the importance weight under the model of all samples in the prior is estimated from single trajectories.
This is approximately the difference between the Boltzmann weight and the weight under the GFN model.
Samples with high importance weights are preferentially chosen for noising and addition to the noised buffer.

\section{Latent Distributions}
\begin{figure}
\centering
\includegraphics[width=\textwidth]{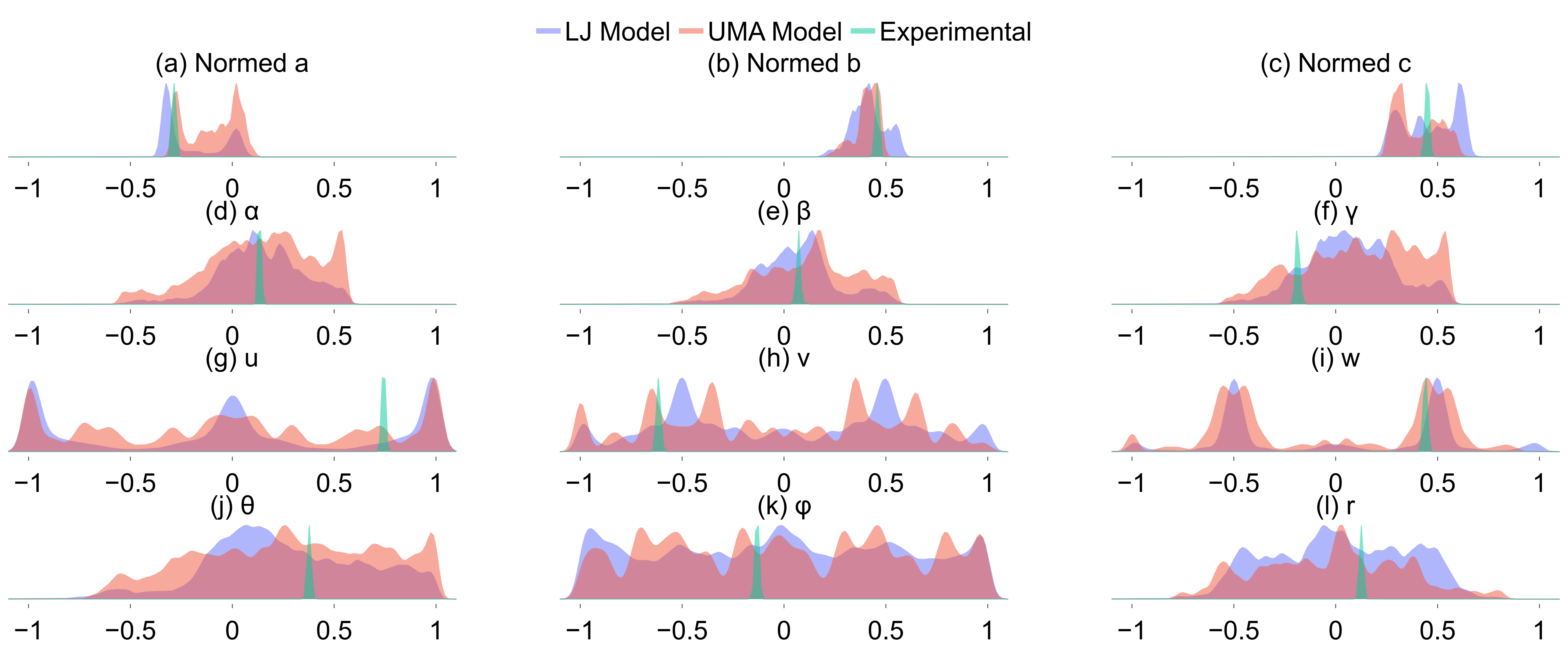}
\caption{\label{fig:mipcas_marginals} Overlays of the 1D marginal latent distributions sampled on MIPCAS in $P\bar{1}$ from the LJ and UMA models. 12 latent dimensions are shown, with location of experimental polymorph highlighted in green.}
\end{figure}

In Figure~\ref{fig:mipcas_marginals}, we visualize the latent space distributions for $P\bar{1}$ MIPCAS sampled from the LJ and UMA models, in the form of 1D marginal distributions over the 12 latent dimensions.
The 66 2D contour plots over the same are given below.

The low-energy crystals under the Lennard-Jones potential arise from the interplay between steric repulsion and dense packing.
The observed distributions, particularly in the asymmetric unit dimensions, are dominated by this consideration.
Common patterns are empirically observed~\cite{groom2016cambridge}, and indeed, theoretically derivable~\cite{tuckerman2023topological,hahn1983international}, for each space group.

In addition, molecule symmetries, combined with the symmetry of the normalizer for each space group, result in sets of identical physical structures (up to rotation, translation) represented by different latent configurations.
This is directly observable in the symmetries of the asymmetric unit centroid distributions.
That the generator learns these complex and high-dimensional symmetries directly from the energy function is a very encouraging signal of model convergence.

\begin{figure}
\centering
\includegraphics[width=\textwidth]{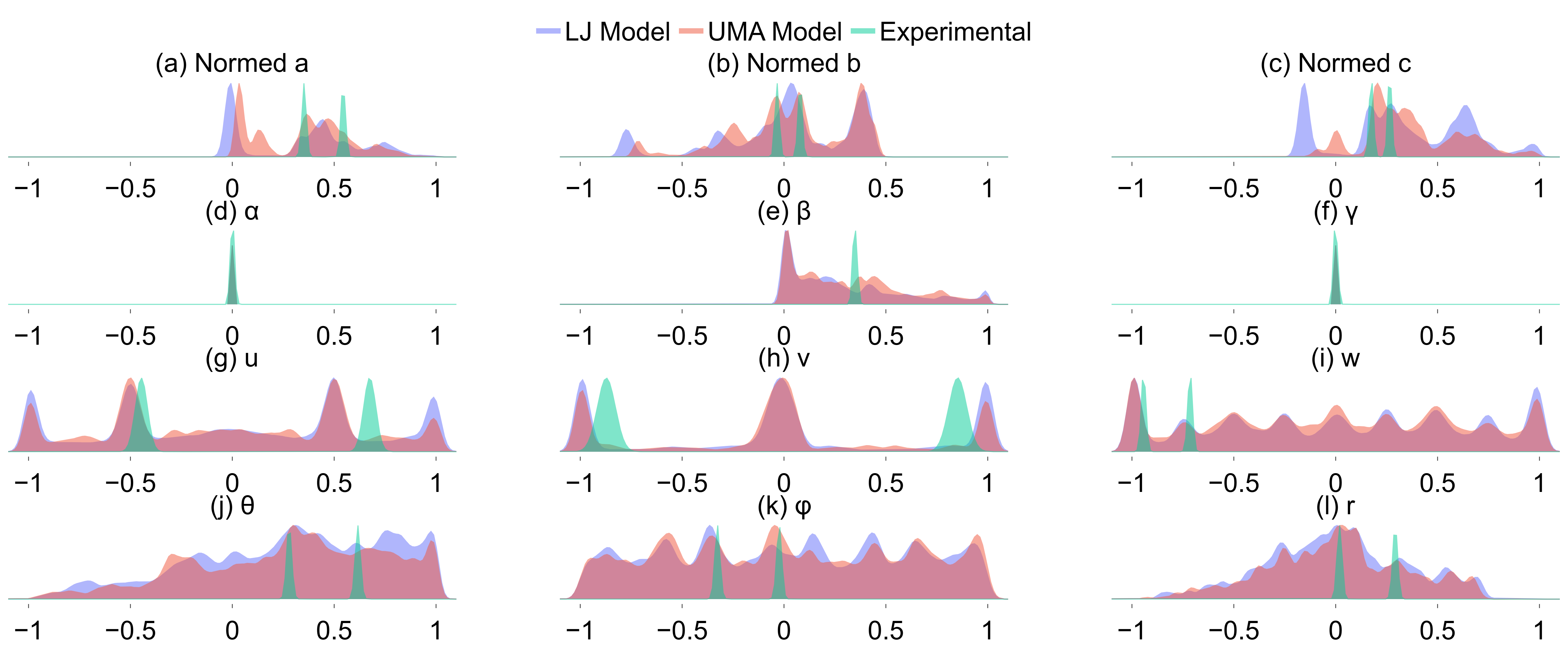}
\caption{\label{fig:nehzor_marginals} Overlays of the 1D marginal latent distributions sampled on NEHZOR in $P2_1/c$ from the LJ and UMA models. 
12 latent dimensions are shown, with locations of experimental polymorphs highlighted in green.}
\end{figure}
In Figure~\ref{fig:nehzor_marginals}, we show the same set of 1D distributions for NEHZOR in $P2_1/c$.
Once again, we observe symmetric patterns in the asymmetric unit positions, though distinct from those in $P\bar{1}$ for MIPCAS due to the difference in molecule and space group.

\begin{figure}
\centering
\includegraphics[width=\textwidth]{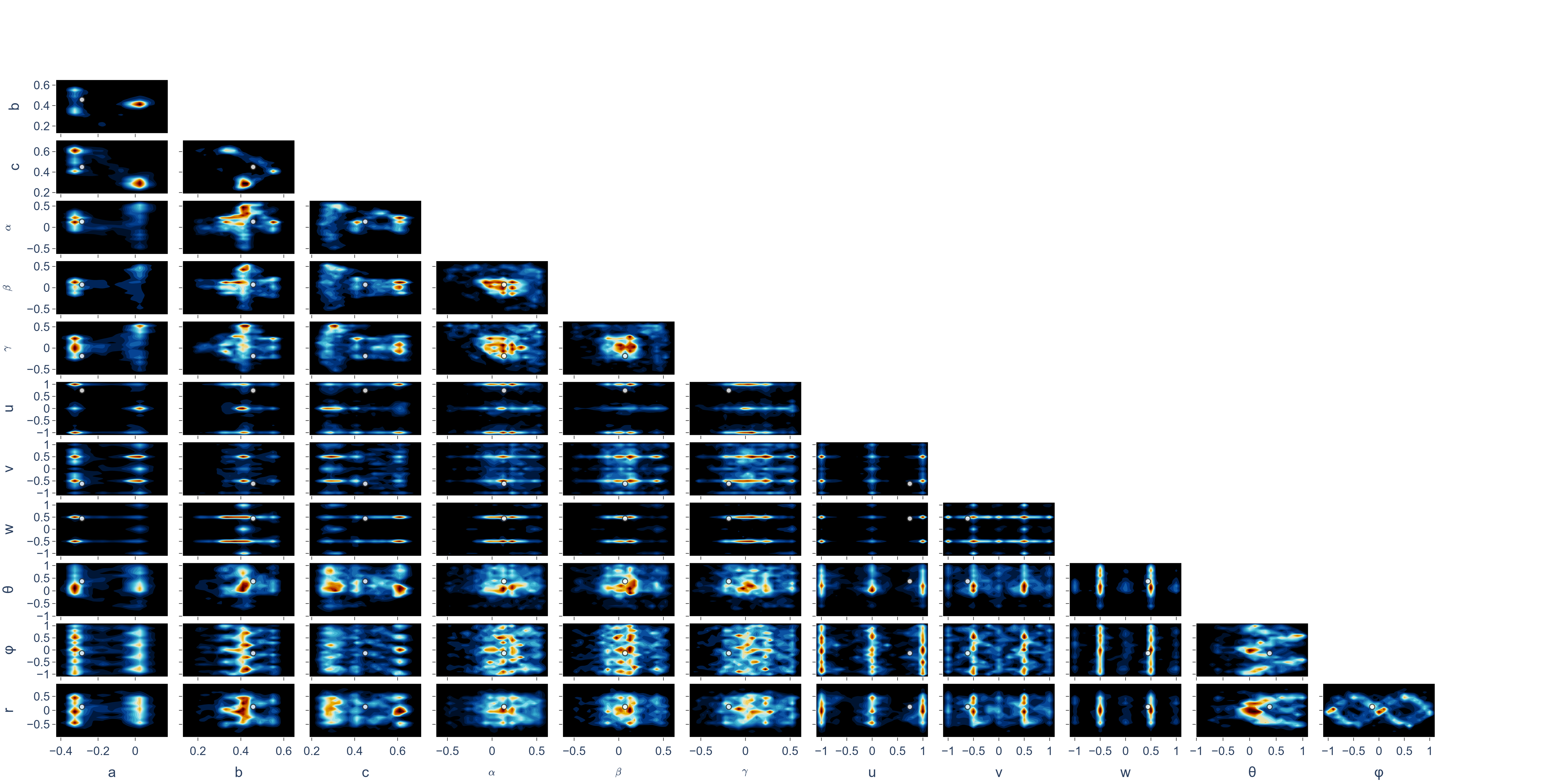}
\caption{\label{fig:staircase1} Staircase plot providing all the 2D marginal distributions over the 12 latent dimensions for 10k samples from the LJ based model for $P\bar{1}$ MIPCAS.}
\end{figure}

\begin{figure}
\centering
\includegraphics[width=\textwidth]{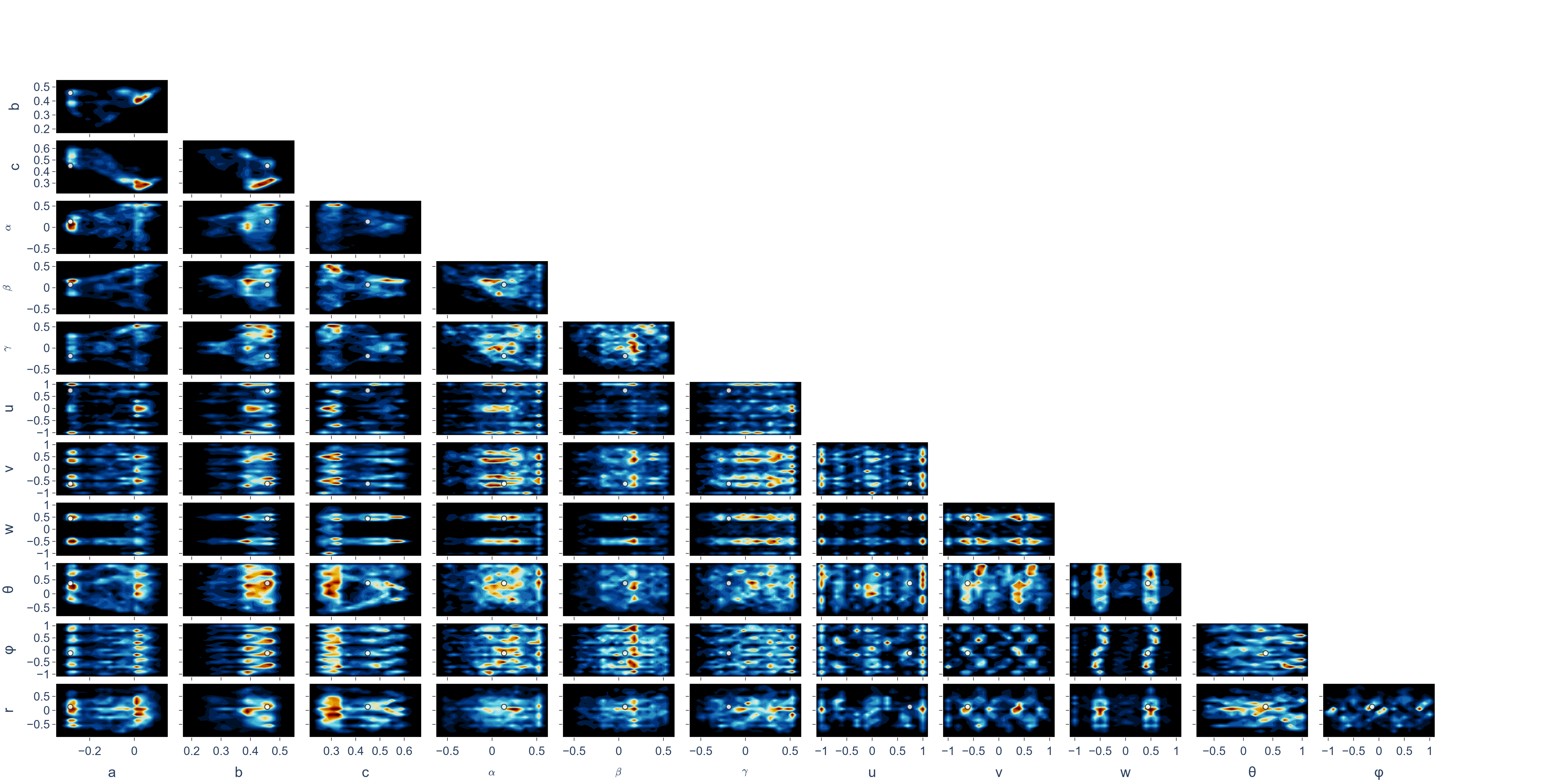}
\caption{\label{fig:staircase2} Staircase plot providing all the 2D marginal distributions over the 12 latent dimensions for 10k samples from the UMA based model for $P\bar{1}$ MIPCAS.}
\end{figure}

\begin{figure}
\centering
\includegraphics[width=\textwidth]{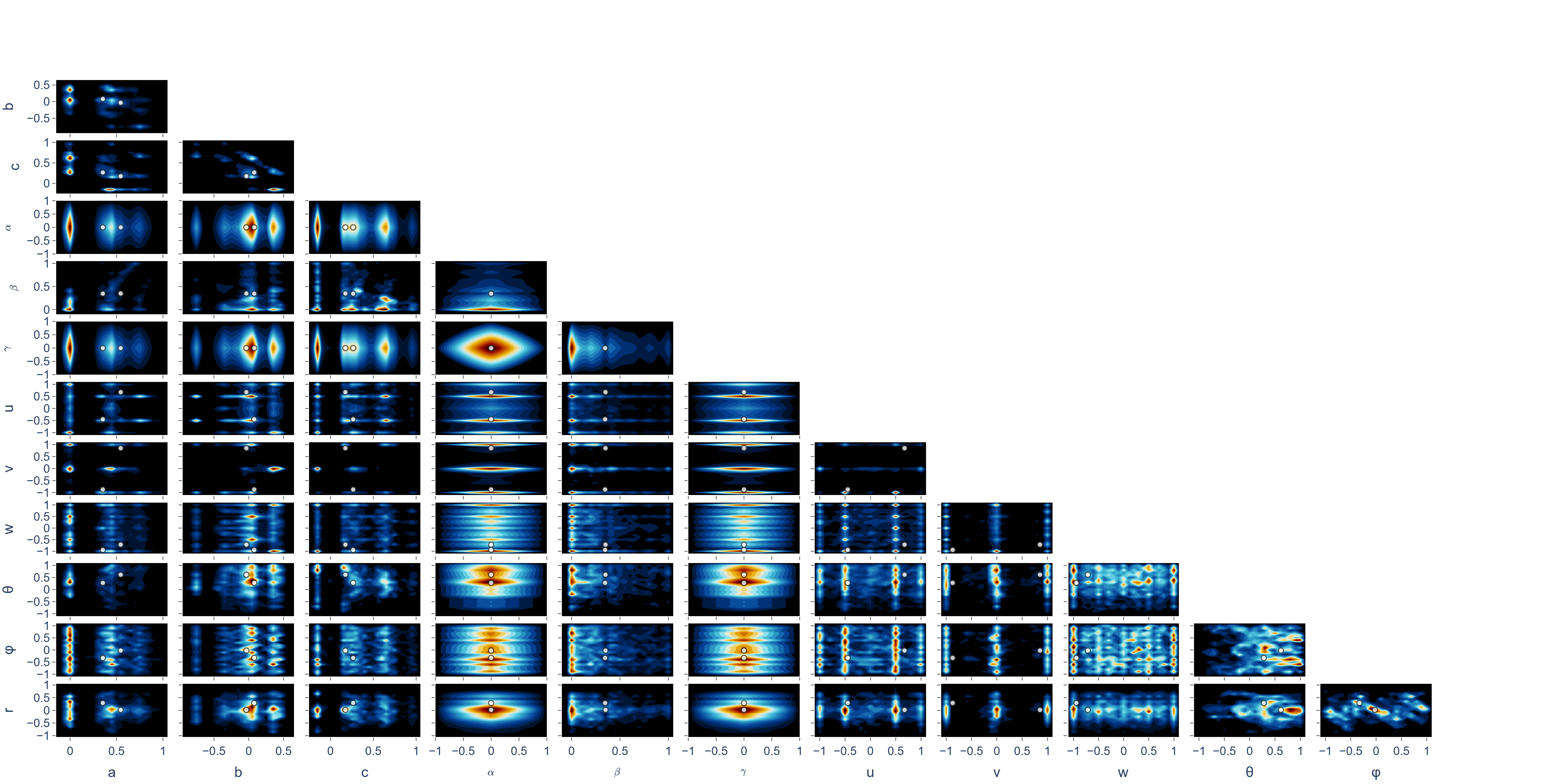}
\caption{\label{fig:staircase3} Staircase plot providing all the 2D marginal distributions over the 12 latent dimensions for 10k samples from the LJ based model for $P2_1/c$ NEHZOR.}
\end{figure}

\begin{figure}
\centering
\includegraphics[width=\textwidth]{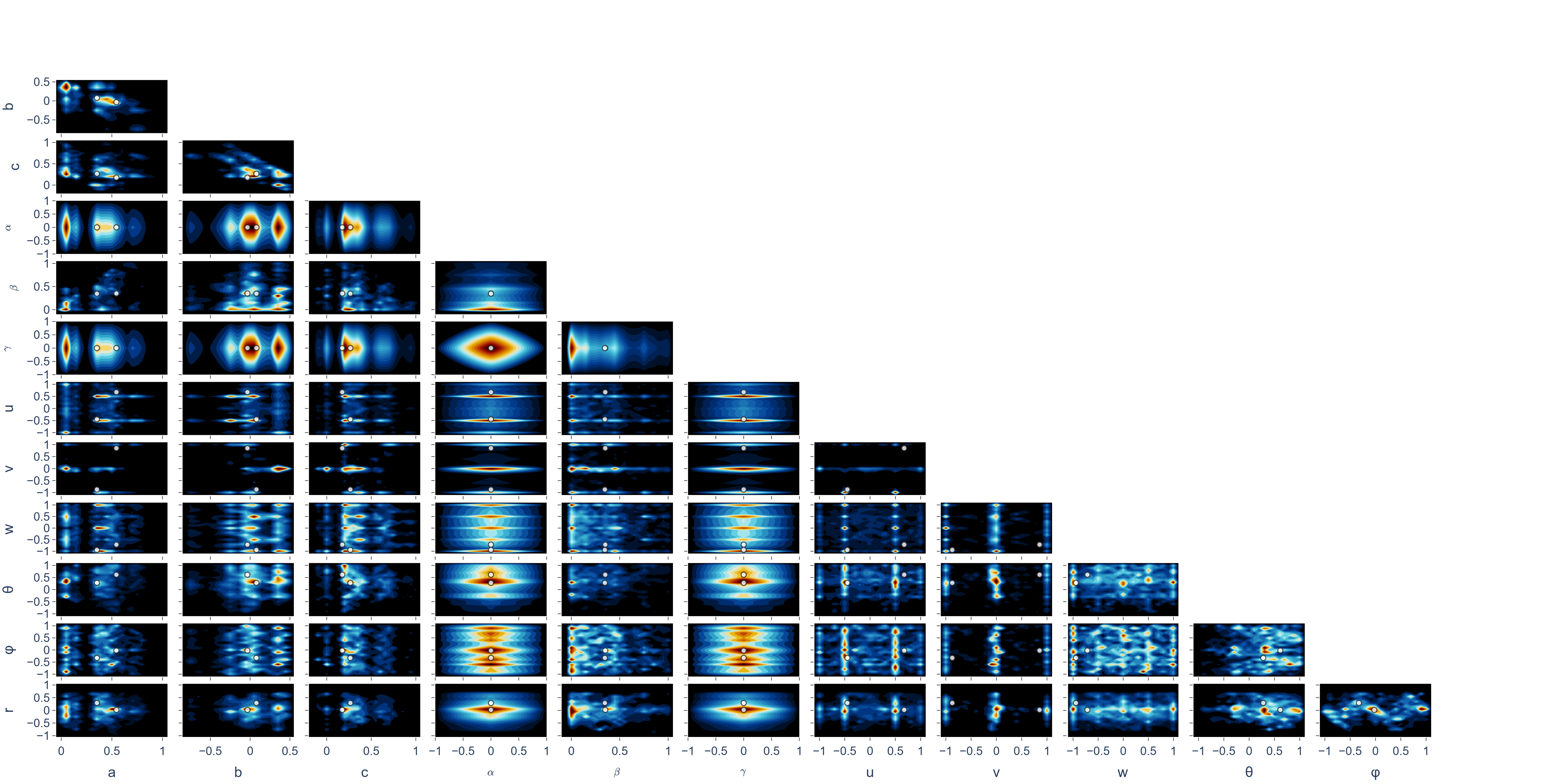}
\caption{\label{fig:staircase4} Staircase plot providing all the 2D marginal distributions over the 12 latent dimensions for 10k samples from the UMA based model for $P2_1/c$ NEHZOR.}
\end{figure}

We share in Figures~\ref{fig:staircase1}-~\ref{fig:staircase4} staircase plots showing all the 2D histograms over the 12 latent dimensions for the sampling distributions of the models trained under LJ and UMA, for $P\bar{1}$ MIPCAS and $P2_1/c$ NEHZOR.

\section{Forward-to-Backward Step Ratio}

At the beginning of Phase 3 of training, the forwards:backwards step ratio is dynamically adjusted to ensure that the slope and intercept of the trajectory balance parity plot (see Figure~\ref{fig:tb_parity}) are below target thresholds.
A regression is fit as $(\log{P_f}+\log{Z_\theta}) = m(\log{P_b}+\log{R}) + b$, with slope $m$ and intercept $b$, and target metrics $e_{slope}=|1-m|$ and $e_{int}=\frac{|b|}{\sigma(\log{P_f}+\log{Z_\theta})}$.
These metrics are updated on a rolling basis, and every 20 steps are used to update the fwd:bwd step ratio.
If the metric $M=max(\frac{e_{slope}}{c_{slope}}, \frac{e_{int}}{c_{int}})$ is greater than 1.2, the fwd:bwd ratio is cut in half.
If $M$ is less than 0.8, the fwd:bwd ratio increases by 5\%.
In our experiments, we enforce high relative accuracy on the prior with relatively tight coefficients $c_{slope}=0.1$ and $c_{int}=1$.
As the forward policy converges, if the same metrics evaluated over the on-policy sample distributions go below the slope or intercept coefficients, the smaller of the forward metrics or pre-set coefficients is used, in order to tighten the convergence criteria. 

\section{Softened Lennard-Jones Potential}

The explosive $\frac{1}{r^{12}}$ repulsive wall in the standard Lennard-Jones potential  makes it unsuitable for guiding our generative model, which does not handle extremely high-energy / low-reward samples well due to the nature of the loss.
Detailed studies have shown that the general, physical intermolecular repulsion is closer to an exponential functional form~\cite{buckingham1938classical,kaplan2006intermolecular}, as captured in the commonly-used Buckingham potential.
The Buckingham potential however has a negative asymptote at $r=0$, which poses a significant danger particularly in early training where crystal structures are poor and atoms frequently overlap.

We define a smooth Lennard-Jones type potential energy function which has exactly the common 12-6 functional form for $r>\sigma$, outside the hard repulsive wall, and exponential below, the exponential Lennard-Jones energy,
\begin{equation}
E^{ij}_{ELJ}(r) =
\begin{cases}
4\big( (\frac{\sigma}{r})^{12}-(\frac{\sigma}{r})^6\big), & r > \sigma, \\
0, & r=\sigma,\\
A e^{-k(r-\sigma)} + B, & r < \sigma, \\
\end{cases}
\end{equation}
for $\sigma$ the sum of van der Waals radii of atoms $i$ and $j$, $A=-\frac{F_0}{k}$, $B=-A$, $F_0=-\frac{24}{\sigma}$, and $k$ a stiffness factor taken as 2.5, yielding a per-pair energy of $\approx100$ at $r=0$.

To translate from arbitrary Lennard-Jones units to a physically meaningful range, we rescale our LJ outputs according to the ratio of the minimum LJ and UMA energies from a sample of their respective prior datasets.
This gives meaning to the temperature the LJ models are trained at, and allows for comparisons between the two distributions.

\section{Auxiliary Energies}
To help guide the generation towards low-energy states, particularly early in training, where the model often veers towards lower densities to avoid severe van der Waals clashes, we apply a penalty for unphysical densities, 
\begin{equation}
    E_\rho=ReLU(-(\log(c_p)-\log(0.55)))^2+2ReLU(c_p-0.95)^2,
\end{equation}
with $c_p$ the crystal packing coefficient, defined as the fraction of cell volume occupied by molecules, and $ReLU$ as the rectified linear unit.
This function penalizes low densities asymptotically towards 0 below packing coefficients of 0.55, and high densities quadratically above 0.95 - the physically reasonable range as observed in experimental crystallography~\cite{groom2016cambridge}.
Inside this range, $E_\rho$ goes to zero.

To constrain the model to the standard latent parameters~\cite{togo2024spglib,sykes2024has}, we also apply penalties to violations of the standardization criteria for each crystal system via quadratic penalties as $E_{reduce}$.
This is distinct from the definition of said crystal system, for example, in the standard setting of a monoclinic crystal the angle $\alpha$ is always definitionally $\frac{\pi}{2}$ and $\beta$ is free.
To ensure the generator learns on the correct sub-manifold for each crystal system, the relevant universal constraints are also applied via quadratic energy penalties.

\begin{itemize}
    \item Triclinic
    In triclinic unit cells, the standard cell is identical to the Niggli reduced cell.
    
    $c \geq b \geq a$

    $\lvert\cos{\alpha}\rvert\leq\frac{b}{2c}$
    
    $\lvert\cos{\beta}\rvert\leq\frac{a}{2c}$
    
    $\lvert\cos{\gamma}\rvert\leq\frac{a}{2b}$

    $ab\cos\gamma+ac\cos\beta+bc\cos\alpha\geq0 $

    \item Monoclinic
    
    $\alpha=\gamma=\frac{\pi}{2}$
    
    $\beta>\frac{\pi}{2}$  

    $|\cos{\beta}|\leq\frac{a}{c}$  
    
\end{itemize}





We apply a general bounding potential to keep samples inside the defined latent space, $(-1,1)^{12}$.
\begin{equation}
E_{bound}(\bm{x}) =
\begin{cases}
\sum_i (x_i+1)^2, & x _i< -1, \\
0, & -1 < x _i< 1, \\
\sum_i (x_i-1)^2, & x _i> 1,
\end{cases}
\end{equation}
for $\bm{x}$ each raw latent space output of the model.

The Jacobian correction terms (derived in Section~\ref{sec:jacob_def}) are converted into effective energies, giving
\begin{equation}
    E_{Jacob}=-k_BT\log{J_{asu}}-k_BT\log{J_{ori}}
\end{equation}
The total energy which goes into the GFlowNet reward function is then,
\begin{equation}
    E_{tot}=E_{phys}+E_\rho+10\cdot E_{reduce}+10\cdot E_{bound} + E_{Jacob}.
\end{equation}

The GFlowNet energy is capable of reaching extremely large values, for example if the generator produces samples with unphysically proximal atoms, or if our boundary conditions are severely violated.
As a result, the reward may span dozens or hundreds of orders of magnitude, and accordingly the TB loss which contains $\log{R(x)}$ may yield destabilizing gradients.
To guard against this we pre-select a reward value below which the reward is softly clipped via logarithm. 
This value is set as $R_{min}=R_{max,d} - R_{range}$, with $R_{max,d}$ the maximum reward in the prior dataset, and $R_{range}=100$ in our experiments.
The clip is effectuated on the sample energy, as $E_{max}=-\frac{R_{min}}{\beta}$ for $\beta$ the inverse sampling temperature.
To avoid saturating physical losses interfering with the signal from the boundary and reduction losses, we softly clip these separately, with the physical energy clipping at $E_{max}$ and the boundary terms at $1.1\cdot E_{max}$.

\section{\label{sec:TB} Model Convergence}

\begin{figure}
\centering
\begin{subfigure}[b]{0.45\textwidth}
    \centering
    \caption{MIPCAS $P\bar{1}$, LJ}
    \includegraphics[width=\textwidth]{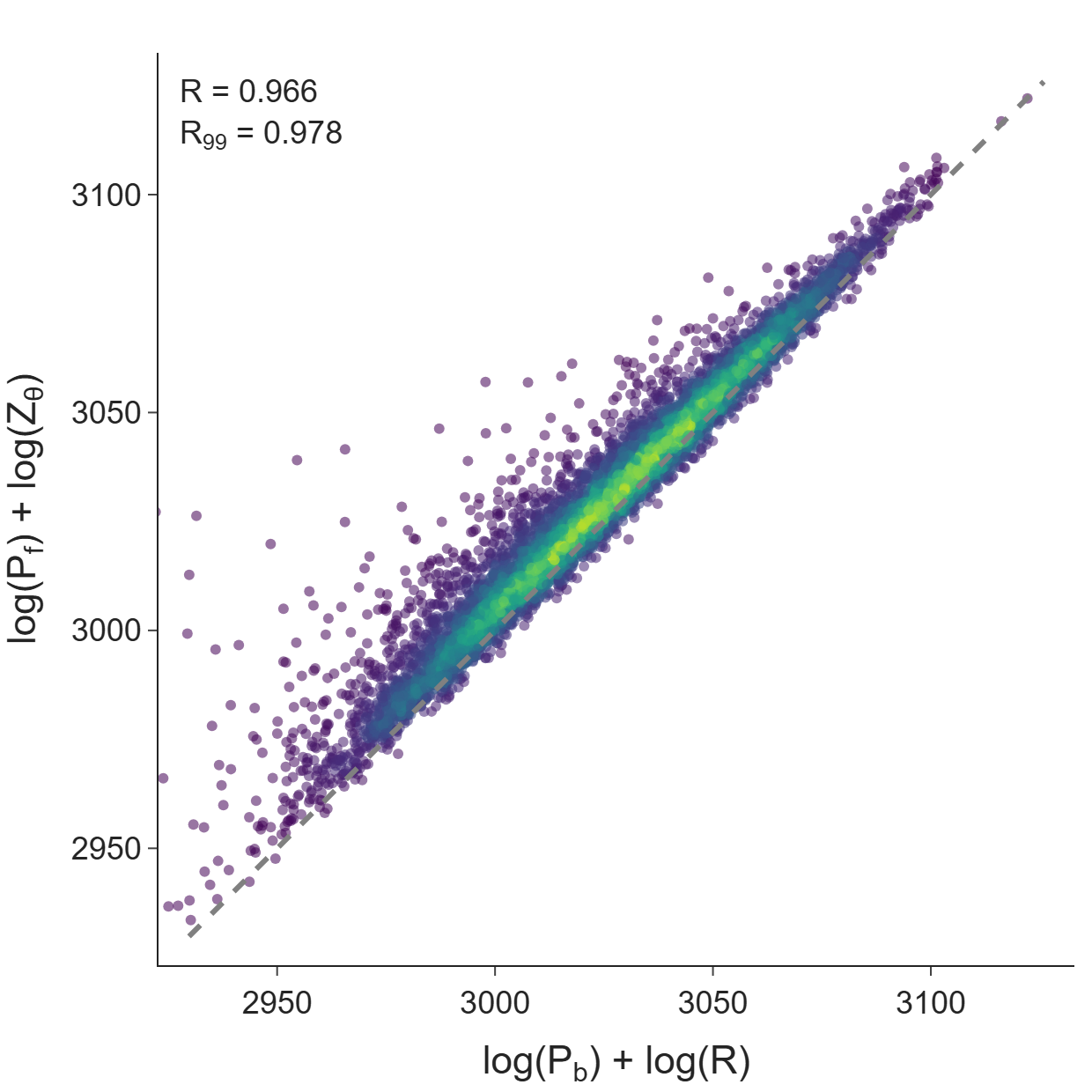}
\end{subfigure}
\hfill
\begin{subfigure}[b]{0.45\textwidth}
    \centering
    \caption{MIPCAS $P\bar{1}$, UMA}
    \includegraphics[width=\textwidth]{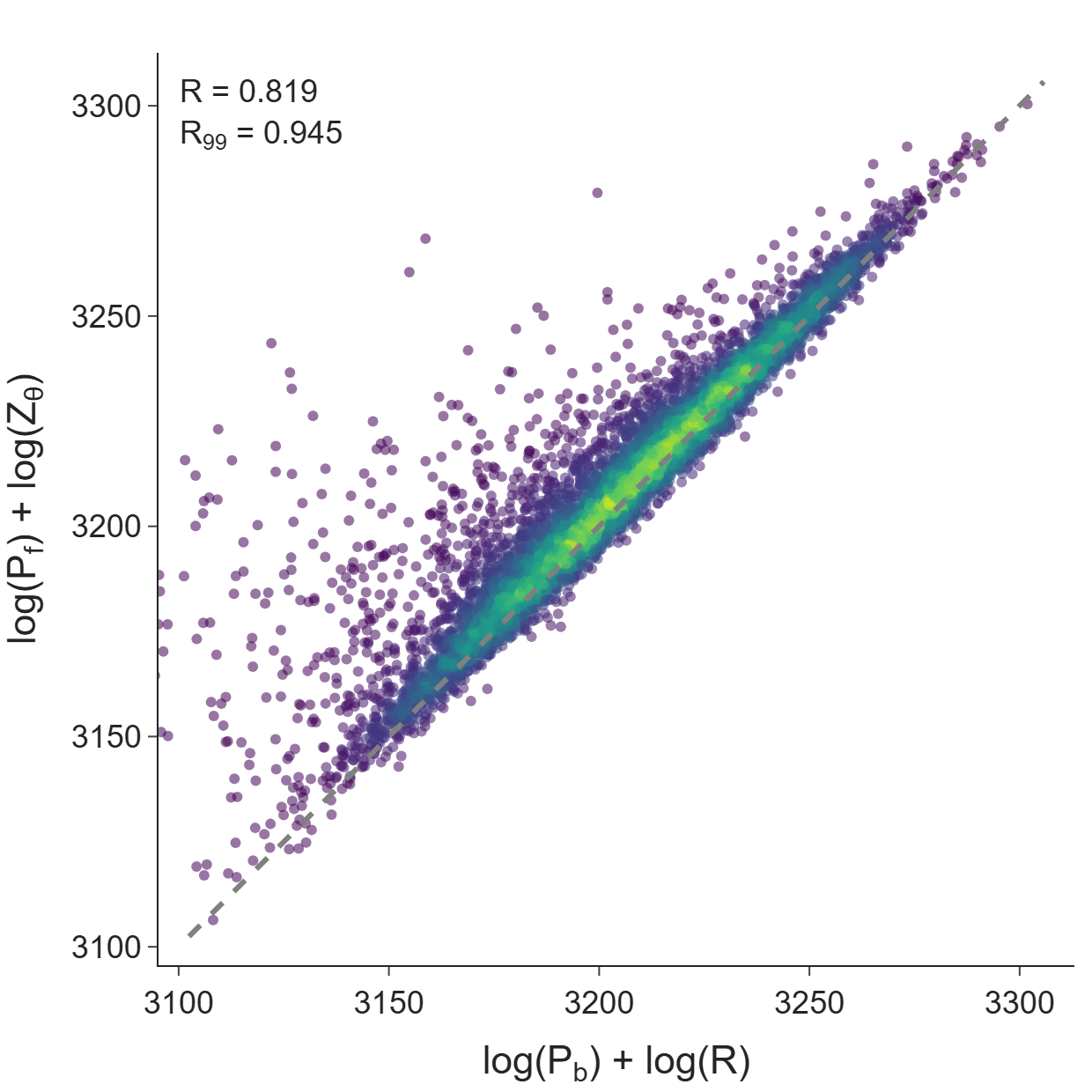}
\end{subfigure}

\vspace{0.5em}

\begin{subfigure}[b]{0.45\textwidth}
    \centering
    \caption{NEHZOR $P2_1/c$, LJ}
    \includegraphics[width=\textwidth]{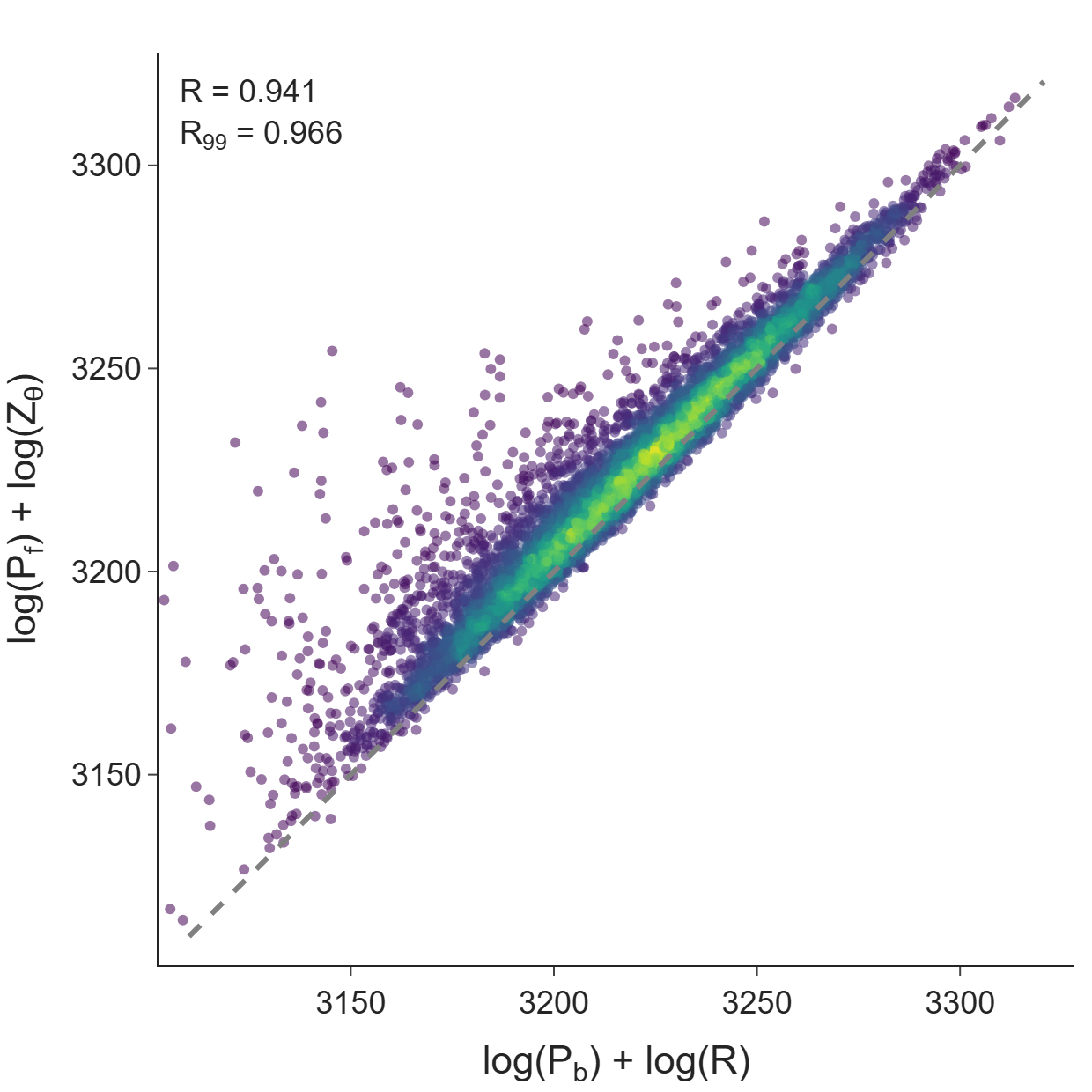}
\end{subfigure}
\hfill
\begin{subfigure}[b]{0.45\textwidth}
    \centering
    \caption{NEHZOR $P2_1/c$, UMA}
    \includegraphics[width=\textwidth]{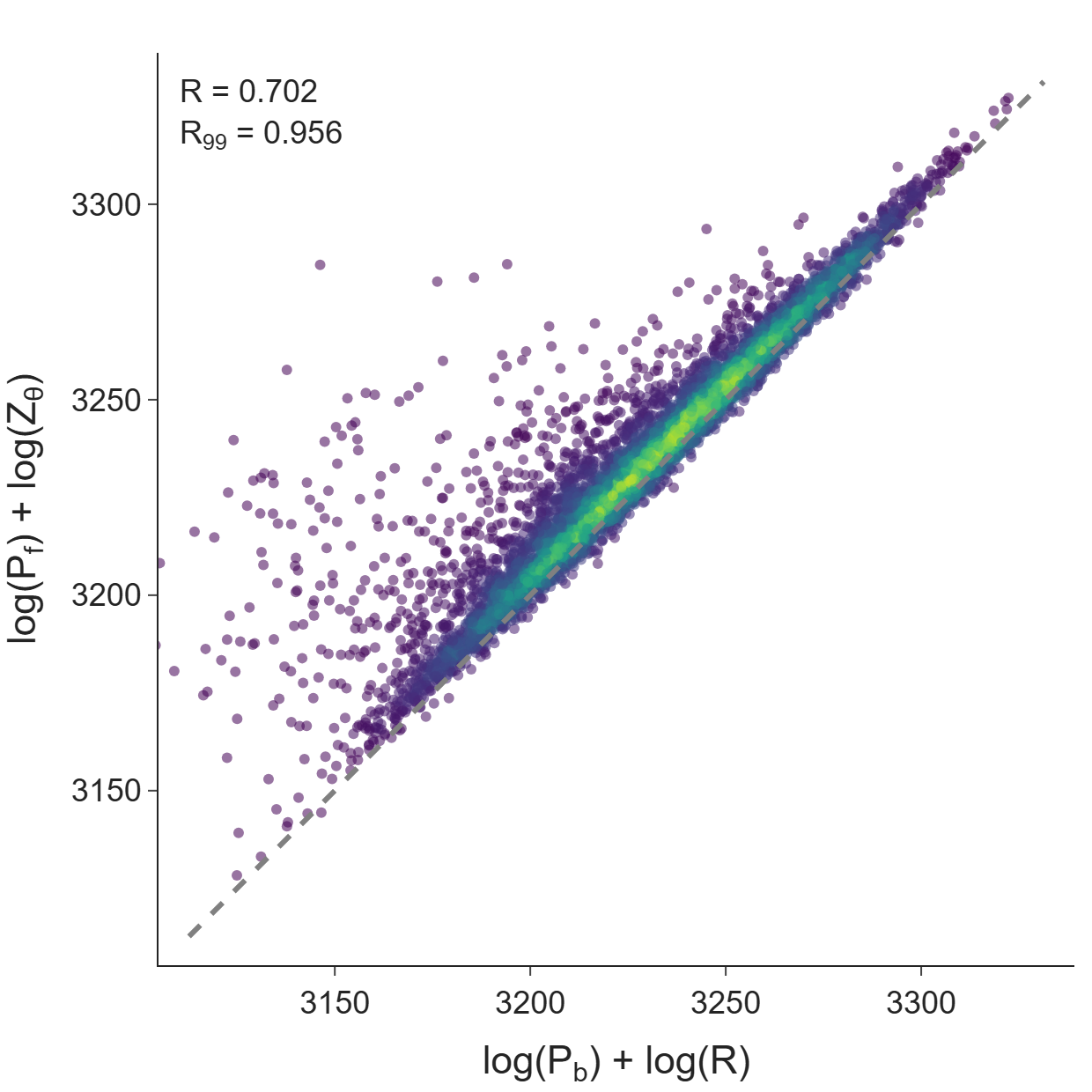}
\end{subfigure}
\caption{Trajectory balance parity plots with full and top 99 percent reward linear fits for 10k samples. Rows: MIPCAS (top), NEHZOR (bottom). Columns: LJ (left), UMA (right).
The x-axes are clipped for readability, leaving off 1-6\% of samples outside of the range, depending on the figure.}
\label{fig:tb_parity}
\end{figure}
We quantify convergence of the on policy TB loss via the Pearson correlation of Equation~\ref{eq:tb_parity}.
These convergence evaluations were done on 10k generated samples, then repeated on the top 99 percent of samples by reward, as just a few very low reward (high-energy) samples can badly skew the metric.
Due to the very high stiffness in the repulsive regime of intermolecular potentials, a relatively small difference in crystal parameters may yield a crystal with very high-energy.
This is one of the key difficulties in converging our models on molecular landscapes.

Accordingly, a sign of incomplete convergence we observe here in Figure~\ref{fig:tb_parity} and in Figure~\ref{fig:energies} is the over-weighting of low reward samples, which appear as samples above the diagonal.
These can be converged out, in principle, given sufficiently long forward training.
By nature, low reward states are rarely sampled, and so their appropriate calibration can take a very long time under our current protocol.
We worked here within a set training budget, and so accept these outliers as errors in the approximation of the distribution. 
As the UMA model works on a more complex surface, we generally observe more difficult terminal convergence, and particularly, residual tails of over-weighted high-energy samples.
From monitoring the effective dimension of the sample distribution during training, we can see that all four GFN models maintained support on their prior distributions and did not undergo mode collapse.

\section{\label{sec:rdf} Radial Distribution Function Earth Mover's Distance}

The physical distances between crystal structures can be estimated via an evaluation of the difference between local `fingerprints', here, in the form of the pairwise intermolecular radial distribution functions between all unique atom pairs.
We extract radial densities for all such pairs on the range of 0-10~\AA~and compute the earth mover's distance between pairs of crystal structures $1$ and $2$.
The distance is computed for atom pair $(i,j)$ with $N$ discrete bins $r$, between normalized radial density functions.
Crystal radial density histograms, $R_m^{i,j}(r)$, with bin width $\Delta r$, are normalized according to
\begin{equation}
    \bar{R}_m^{i,j}(r)=\frac{R_m^{i,j}(r)}{\sum_r{R}_m^{i,j}(r) + \epsilon},
\end{equation}
for $\epsilon$ a small stability factor.
1D pairwise earth mover's distances are then computed as
\begin{equation}
    EMD_{1,2}^{i,j} = \Delta r \sum_{k=0}^N\Bigg| \sum_{r=0}^k\bar{R}_1^{i,j}(r) - \sum_{r=0}^k\bar{R}_2^{i,j}(r) \Bigg|,
\end{equation}
with $N$ the total number of radial bins.
The crystal-crystal RDF EMD is then aggregated by averaging over atom pairs
\begin{equation}
    EMD_{1,2}=\frac{1}{\mathcal{A}}\sum_{i,j \in \mathcal{A}}EMD_{1,2}^{i,j},
\end{equation}
for $\mathcal{A}$ the set of atom pairs with non-vanishing density, corresponding to the elements present in either crystal. 

Since in this work, we are taking radial distributions over all unique atom pairs, rather than all pairs of elements or atom types, the earth mover's distance for a given atom pair is the distance and quantity of radial density that must physically change between two configurations.

\section{\label{sec:jacob_def} Jacobian}
Our base distribution is defined by (1) the CSD statistics for the box degrees of freedom $(a,b,c,\alpha,\beta,\gamma)$, (2) a uniform distribution over molecule centroid locations in T(3), and (3) a uniform distribution over rotations in SO(3). Our latent space is uniform in asymmetric unit positions as well as over the spherical coordinates, and so must be corrected.

For the transformation from the fractional basis of the crystal asymmetric unit to the cartesian coordinates, the Jacobian correction accounts for the difference in the configuration space volume as a function of the unit cell size, as compared to the volume in fractional coordinates, which is constant.
The relative difference in volume scales as $V_{asu}=V_{cell}/Z$, for $Z$ the crystal symmetry multiplicity, and the appropriate weight factor is
\begin{equation}
\label{eq:jacob1}
    J_{asu}=\left(\frac{V}{Z}\right),
\end{equation}
where $V$, and $Z$ are the unit cell volume and symmetry multiplicity, respectively, effectively penalizing smaller cells with accordingly smaller phase space volumes.
This can be derived from the change of variables between fractional coordinates of the asymmetric unit and cartesian coordinates, 
\begin{equation}
\begin{split}
    dx\ dy\ dz&=\left|\det\frac{\partial(x, y, z)}{\partial(u, v, w)}\right|\ du\ dv\ dw,\\
    &=|\det \hat{H}_{asu}|\ du\ dv\ dw,\\
    &=V_{asu}\  du\ dv\ dw,
\end{split}
\end{equation}
where $\hat{H}_{asu}$ is the box matrix of the asymmetric unit, and $V_{asu} = \frac{V}{Z}$ is its determinant, the volume of the asymmetric unit.

The correction for orientation degrees of freedom comes from the change of variables from the spherical axis-angle coordinates to the space of unit quaternions. 
For $r$ the rotation angle and $(\theta, \phi)$ the polar and azimuthal angles, the result, modulo constants, is the Haar measure on SO(3)~\cite{yershova2010generating}
\begin{equation}
    d\mu_{Haar}\propto\sin^2{(r/2)}\cdot\sin{\theta}\ dr \ d\theta\ d\phi.
\end{equation}
This leaves us with the correction term $J_{ori}=\sin^2{(r/2)}\cdot\sin{\theta}$.

\end{document}